\documentclass[a4paper]{spie}
\usepackage{amsmath,amsfonts,amssymb}
\usepackage{graphicx}
\usepackage[colorlinks=true, allcolors=blue]{hyperref}
\usepackage[english]{babel}

\usepackage{subcaption}
\usepackage{authblk}
\usepackage{geometry}
\usepackage{tikz}
\usetikzlibrary{quotes,angles,calc,patterns,shapes.geometric}
\usepackage{soul}
\usepackage{multirow}
\usepackage{titlesec}

\usepackage{textcomp, gensymb}
\usepackage{csquotes}
\title{Transmission curves of narrow-band filters in large-FoV and fast astronomical instruments}

\author[a]{Federico Battaini}
\author[a,b]{Roberto Ragazzoni}
\author[a,b]{Antonino P.\,Milone}
\author[a]{Gabriele Cremonese}

\affil[a]{Istituto Nazionale di Astrofisica - Osservatorio Astronomico di Padova,  Vicolo dell'Osservatorio 5, 35122 Padova,  Italy}
\affil[b]{Dipartimento di Fisica e Astronomia ``Galileo Galilei'', Universit{\'a} di Padova, Vicolo dell'Osservatorio 3, 35122 Padova,  Italy}
\authorinfo{Send correspondence to: federico.battaini@inaf.it}

\begin{document}
\maketitle

\begin{abstract}
%, in particular, in recent years, to identify chemical abundances in globular clusters stellar populations. 
%APM: Non e' corretto dire che servono ad identificare le singole righe (nel nostro caso ad esempio identifichiamo l'effetto di --molte-- bande molecolari sui colori e magnitudini delle stelle. Comunque non si identificano mai le righe. Ma l'effetto sul flusso di una variazione di abbondanza (o di parametri atmosferici).
%Direi forse (anche se non proprio completo, e' comunque corretto):
Narrow-band filters are often used to %identify single spectral lines 
 constrain the chemical composition of astronomical objects through photometry.
%APM. Qui aggiungerei una mezza frase per collegare le due cose: % OK FB
A challenge to derive accurate photometry is that narrow-band filters are based on interference of multiple reflections and refractions between thin layers of transparent dielectric material. When the light rays reach the surface of a filter not perpendicular to it, they cross the layers obliquely travelling a path longer than the thickness of the layers and different for each inclination.
 %APM
 % la frase sopra mi suona colloquiale/italiana... forse qualcosa tipo: %% CAMBIATA
 % When the inclination is not perpendicular to the filter surface, the rays can cross layers (specifica di cosa) with different depth ?
 This %reveals itself 
  results 
  in a blue-shift of the central wavelength and 
   a
  distortion of the transmission curve. 
 Hence, particular care should be taken when narrow band filters are used in presence of small f-numbers and large non-telecentric angles, as frequent in the large field of view (FoV) instruments.
 %APM: specifica a quale effetto ti riferisci
Sometimes, the broadening and central wavelength shift of the transmission curve are considered and compensated in the design of filters for instruments with a small f-number.
 %APM: questa frase e' incompleta ?  Oppure intendiamo dire che al contrario da quanto fanno altri cosideriamo... ? La seconda, ma ho riformulato FB
Here we consider the combined effect of small f-number, non-telecentricity and large FoV. 
Where single spectral lines are considered, a shift in central wavelength or a change in the shape of the transmission curve may introduce an instrumental dispersion in luminosity and in the linked color indices. 
We found that transmission curves of narrow band filters can be significantly different in shape than the nominal ones.
The bottom limits for filters' effective FWHM for each f-number; the monotonic behavior of the blue-shift with distance from the center of FoV; the monotonic quality decrease of the transmission curves and the photometric dispersion introduced by the filters are computationally estimated.  This work could represent a useful tool to evaluate the fitness of a particular filter at a particular facility.
\end{abstract}

\keywords{Thin-layer filters, Large Field of View Instruments, Photometry, Transmission curve distortion, Stellar Populations}

\section{INTRODUCTION}
Uses of narrow-band filters on large field-of-view (FoV) and fast instruments may represent a challenge in photometry, in particular for abundances estimation. Non-perpendicularity (we will call this non-normality so on) to the filter surface of the incoming rays translates into a blue shift of the central wavelength of the narrow-band filters based on thin-layers interference.  The effect is sometimes taken into account to compensate for the broadening and shifting of the transmission curve for instruments with a small F-number. While the combined effect of small F-number, non-telecentricity, and large FoV needs to be considered in challenging optical configurations. We considered the various factors to define bottom limits to filters FWHM and to evaluate the photometric dispersion introduced by the filters along with the distance from the center of the FoV. 
%In the section \ref{sec:thinlayer} we will recollect some useful results about the narrow-band filters in a converging beam as the equations \ref{eq:Lissb1} and in the section \ref{sec:ntaspace} we define the mathematical space of the non-telecentric angles, the space over which we will perform our numerical analysis based on equation \ref{math:lambdatheta}. In the appendix we present the basics on interference filters starting from a Fabry-Perot interferometer. In the sections \ref{sec:fnumeberres} and \ref{sec:ntaresponse} we present the results of our numerical analysis of the transmission curve response under different conditions. 

\section{THIN-LAYERS FILTERS IN A CONVERGING BEAM}\label{sec:thinlayer}
%%%%%%%%%%%%%%%%%%%%%%%%%%%%%%%%%%%%%%%%%%%%%%%%%%%%%%%%%%%%%%%%%%%%%%%%%%%%%%%%%%%%%%%%%%%%%%%%%%%%%%%%%%%%%%%%%%%%%%
%%%%%%%%%%%%%%%%%%%%%%%%%%%%%%%%%%%%%%%%%%%%%%%%  CONVERGING BEAM %%%%%%%%%%%%%%%%%%%%%%%%%%%%%%%%%%%%%%%%%%%%%5%%%%%%
%%%%%%%%%%%%%%%%%%%%%%%%%%%%%%%%%%%%%%%%%%%%%%%%%%%%%%%%%%%%%%%%%%%%%%%%%%%%%%%%%%%%%%%%%%%%%%%%%%%%%%%%%%%%%%%%%%%%%%
In astronomical optical instruments, filters are frequently placed in a filter wheel near the focal plane, thus in a converging beam zone. A parallel beam of e.m. radiation from a single point-like astronomical source, which thus arrives on the telescope at a single angle, in the convergent beam zone is transformed into a cone that scans all angles between that of the chief ray and an extreme one, of the marginal rays, determined by the focal ratio. By depicting various small sub-apertures of the entrance pupil we can see that each leaves its own footprint on the filter and arrives at different angles depending on its position in the pupil. \\
Because of the blue shift of the central wavelength when a filter is illuminated by non-perpendicular light (see \ref{appen:optical}) each of these sub-apertures is then filtered slightly differently. Putting the size of these sub-apertures to zero, we can say that each \textit{ray}, out of multiple coming from the same source, finds a filter with a different transmission curve in front of it. In the focal plane, we shall then find the radiation resulting from the sum, i.e., the integral, on each point of the entire pupil: we shall thus have the sum of infinitesimally slightly different filters, the result of which will be in each case, as much or as little, different from the nominal filter whose wavelength transmission curve is given for a parallel beam. \\

In Lissberger\cite{Lissberger:59II} is found a first analytical treatment in a telecentric configuration. 
In the approximation of small angles and invariance of the transmission curve, a converging beam of semi-aperture angle $\alpha$, the shift is half that found for a parallel beam hitting the filter with angle $\alpha$. Furthermore, the full width half maximum (FWHM) of the passband results: 
\begin{equation}\label{eq:Lissb1}
    FWHM \propto \frac{\alpha^2}{\mu_*^2}
\end{equation}
%under the same assumptions, to be proportional to the square of the beam half-opening angle and inversely to the square of the thin-layer refractive indices.\

Lissberger and Wilcock \cite{Lissberger:59II} conclude that for F numbers not too \textit{fast}, greater than $F/3.3$ in their choice of parameters\footnote{$\lambda_0=5000$\AA, $n_*=2$ zinc sulfide spacer, $FWHM_0=50$\AA, and refractive index of zinc sulfide at $5000$\AA $\mu_z=2.39$}, the deterioration of filter performance is negligible. Even for smaller F numbers, the deterioration remains minor if the shift of the central wavelength is taken into account when designing the filter. \\
%They find, for example, that to achieve maximum transmission at $\lambda=5000\AA$ for a focal ratio $F/2.0$ one must construct a filter with $\lambda_0=5020\AA$. They also conclude that deterioration is greater for narrower filters. \\
%As we will see in the \ref{Chap:Numerical} section, each of these results is confirmed by the numerical simulations and qualitative evaluations with which we will expand on what is expressed in this section. \\
The limitations of an analytical method lie mainly in the difficulty of dealing with transmission functions that are not easily integrable analytical functions; they didn't take into account the presence of the non-telecentric angles (NTAs). \ For sources not positioned at the center of the field of view the axis of the converging cone (its chief ray) is no longer normal to the filter. It is not sufficient to take the passband shift into account at the design stage to neutralize filter deterioration, since the presence of non-telecentric angles results in differential variations within the field of view: sources at different distances from the center of the field are therefore filtered differently. \\
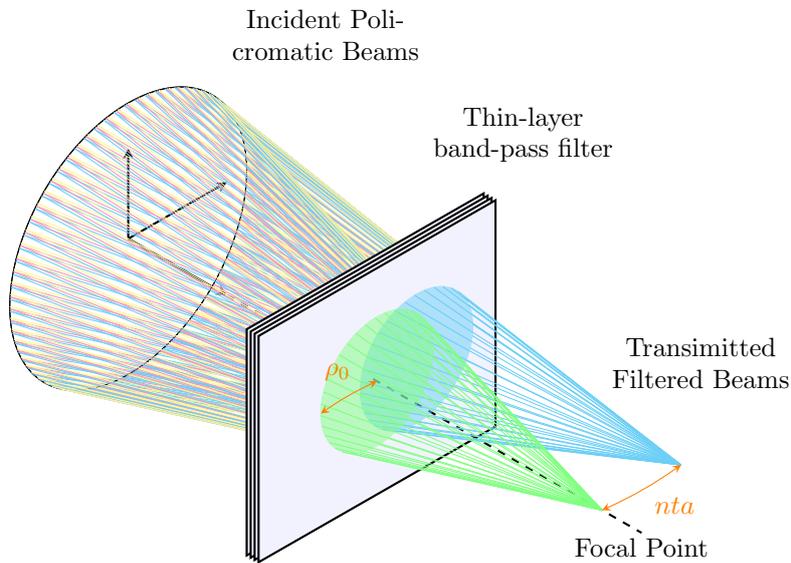
\begin{figure}\label{figure:cone}
	\centering
	% Interference filter layout
% Author: Battaini Federico
% Based on Polarizing Microscope by Cyril Langlois
% This TikZ code sketches the light behavior during its travel in an inteference filter

\begin{tikzpicture}[x={(0.866cm,-0.5cm)}, y={(0.866cm,0.5cm)}, z={(0cm,1cm)}, scale=0.6,
    %Option for nice arrows
    >=stealth, %
    inner sep=0pt, outer sep=2pt,%
    axis/.style={thick,->},
    wave/.style={thick,color=#1,smooth},
    polaroid/.style={fill=black!60!white, opacity=0.3},
]
    % Colors
    \colorlet{darkgreen}{green!50!black}
    \colorlet{lightgreen}{green!80!black}
    \colorlet{darkred}{red!50!black}
    \colorlet{lightred}{red!80!black}
    \colorlet{lightblue}{blue!80!black}

    % Frame
    \coordinate (O) at (0, 0, 0);
    \draw[axis] (O) -- +(2.5, 0,   0);% node [bottom] {Direction of Propagation};
    \draw[axis] (O) -- +(0,  2.5, 0);% node [right] {y};
    \draw[axis] (O) -- +(0,  0,   2);% node [above] {z};

    \draw[thick, dashed] (2.5,0,0) --(6,0,0);

	%CONE II BEFORE FILTER	
	\draw plot[variable=\y,domain=0:360,samples=180] (0,{3*cos(\y)},{3*sin(\y)});
	\draw plot[variable=\y,domain=0:360,samples=180] (6,{1.5*cos(\y)},{1.5*sin(\y)});
	\foreach \y in {0,5,...,360}{
		\draw[magenta!50] (0,{3*cos(\y+1)},{3*sin(\y+1)}) -- (6.3,{1.425*cos(\y+1)+1},{1.425*sin(\y+1)});
		\draw[cyan!50] (0,{3*cos(\y+2)},{3*sin(\y+2)}) -- (6.3,{1.425*cos(\y+2)+1},{1.425*sin(\y+2)});
		\draw[yellow!50] (0,{3*cos(\y)},{3*sin(\y)}) -- (6.3,{1.425*cos(\y)+1},{1.425*sin(\y)});
		%\draw[yellow!50] (0,{3*cos(\y+3)},{3*sin(\y+3)}) -- (6.3,{1.425*cos(\y+3)+1},{1.425*sin(\y+3)});
	}

    %CONE I BEFORE FILTER
	\draw plot[variable=\y,domain=0:360,samples=180] (0,{3*cos(\y)},{3*sin(\y)});
	\draw plot[variable=\y,domain=0:360,samples=180] (6,{1.5*cos(\y)},{1.5*sin(\y)});
	\foreach \y in {0,5,...,360}{
		\draw[magenta!50] (0,{3*cos(\y+1)},{3*sin(\y+1)}) -- (6,{1.5*cos(\y+1)},{1.5*sin(\y+1)});
		\draw[cyan!50] (0,{3*cos(\y+2)},{3*sin(\y+2)}) -- (6,{1.5*cos(\y+2)},{1.5*sin(\y+2)});
		\draw[yellow!50] (0,{3*cos(\y)},{3*sin(\y)}) -- (6,{1.5*cos(\y)},{1.5*sin(\y)});}
		%\draw[yellow!50] (0,{3*cos(\y+3)},{3*sin(\y+3)}) -- (6,{1.5*cos(\y+3)},{1.5*sin(\y+3)});

    \draw (2.5,2.5,4.5) node [text width=2.5cm, text centered]{Incident Policromatic Beams};
    \draw (11,3.5,1) node [text width=2.5cm, text centered]{Transimitted Filtered Beams};
    
    %FILTRO
    \begin{scope}[thick]
    \foreach \shi in{0,0.1,0.2,0.3}
		\filldraw[fill=blue!5] ({6+\shi},-3,-2.5) -- (6+\shi,-3,2.5) -- ({6+\shi}, 3, 2.5) -- ({6+\shi},3, -2.5) -- cycle;% Faces  
	 \end{scope}	
	 \draw (6.2,3.8,3.5) node [text width=2.5cm, text centered]{Thin-layer band-pass filter};

	\draw[thick,dashed] (6.2,0,0) -- (13,0,0) node[below]{Focal Point};
	\filldraw[cyan!50,opacity=0.5] (6.3,1,0) circle [x={(0,1,0)},y={(0,0,1))},radius=1.425];	
	\foreach \y in {0,10,...,360}
		\draw[cyan!50] (6.3,{1.425*cos(\y+2)+1},{1.425*sin(\y+2)}) -- (12,2,0);
	
	\filldraw[green!50,opacity=0.5] (6.3,0,0) circle [x={(0,1,0)},y={(0,0,1))},radius=1.425];	
	\foreach \y in {0,10,...,360}
		\draw[green!50] (6.3,{1.425*cos(\y+2)},{1.425*sin(\y+2)}) -- (12,0,0);

	\coordinate (a) at (6.3,-1.425,0); \coordinate (b) at (12,0,0); \coordinate (c) at (6.3,0,0); 
	\draw[orange,<->] (a) to[bend left=5] (c);
	\node at (6.3,-1.,0.4) [above, orange]{$\rho_{0}$} ;
	
	\draw[orange,<->] (12,0,0) to[bend right=5] (12,2,0);
	\node at (12.2,1,0) [below right,orange]{$nta$};
\end{tikzpicture}
	\caption{A parallel beam of light is transformed into a convergent beam by the telescope. "Cones" of light from stars at different positions in the field of view present a chief ray that etches the filter at different angles, except in the case of a telecentric system.}
\end{figure}

%Si potrebbe supporre che per eliminare i problemi relativi al fascio convergente e all'angolo non-telecentrico sia quella di posizionare i filtri in una zona di fascio parallelo, magari in un reimaging della pupilla, cosa che peraltro avviene in alcuni strumenti, questo però non è affatto una soluzione. \\ 
%L'invarianza di Lagrange ci dice infatti che il prodotto fra inclinazione e diametro del fascio è costante per cui $$\Theta D=\Theta' D'$$ ad un diametro più piccolo corrisponde un angolo più grande. Questo si traduce in un maggior spostamento della lunghezza d'onda centrale della curva di trasmissione del filtro, e differente per ogni sorgente, stante la necessità di costruire filtri molto più piccoli dell'apertura del telescopio\footnote{Non sarebbe così e fosse possibile costruire un filtro delle dimensioni dell'apertura del telescopio, per cui gli angoli di ingresso anche con un campo di vista grande rimangono tanto piccoli da non costituire un problema di deterioramento delle performance del filtro }.
\subsection{FoV and NTA}
To know the relationship between the position in the FoV of a source and the NTA is simple as long as we know some basic characteristics of the optical instrument: the focal ratio and the position of the exit pupil.

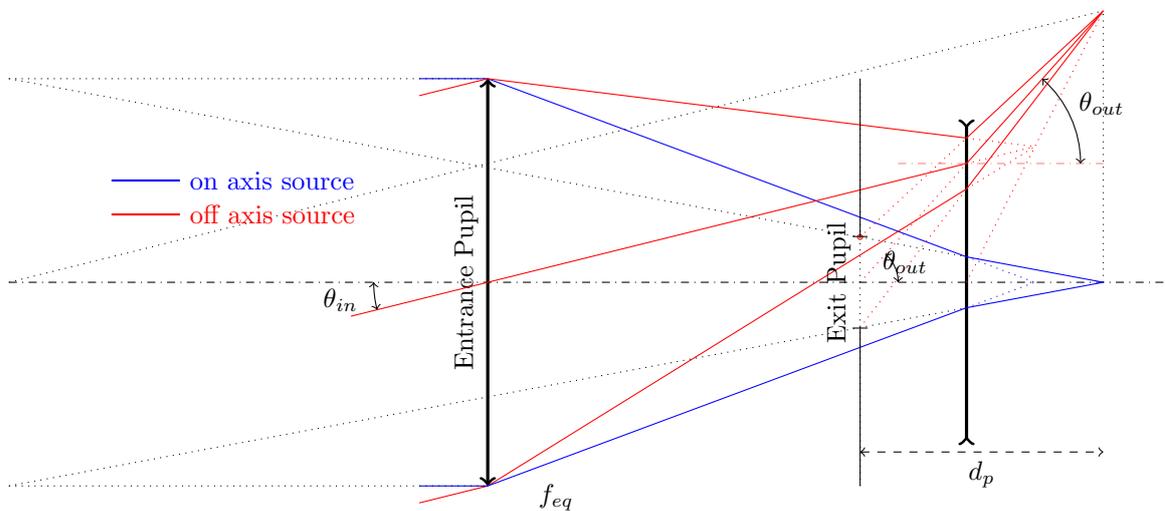
\begin{figure}
	\centering
	\begin{tikzpicture}[scale=0.9]
		\draw[<->, very thick] (0,-3) to (0,3);% node[above left, midway]{Entrance Pupil}; %prima lente
		\draw (0,0) node[above, rotate=90]{Entrance Pupil};
		\draw[>-<, very thick] (7,-2.4) to (7,2.4); %seconda lente
		
		\draw[dash dot] (-7,0) to (10,0); %asse ottico
		
		%% ON AXIS
		\draw[blue] (-1,3) to (0,3) to (7,0.375);
		\draw[dotted] (-7,3) to (0,3);
		\draw[dotted,blue] (7,0.375) to (8,0);
		%% OFF AXIS
		\draw[red] (-1,2.75) to (0,3) to (7,2.125);
		\draw[dotted,red] (7,2.125) to (8,2);
		\draw[red] (7,2.125) to (9,4);
		
		\draw[dotted] (-7,0) to (9,4);
		\draw[dotted] (9,0) to (9,4) ;
		
		\draw[red] (-2,-0.5) to (7,1.75);
		\draw[dotted,red] (7,1.75) to (8,2); %chief red
		\draw[red] (7,1.75) to (9,4);
		
		\draw[dash dot,red!50] (6,1.75) to (9,1.75);
		
		\draw[red] (-1,-3.25) to (0,-3) to (7,1.375);
		\draw[dotted,red] (7,1.375) to (8,2);
		\draw[red] (7,1.375) to (9,4);
		
		\draw[dotted,red] (7,0) to (9,4);
		
		\draw[blue] (-1,-3) to (0,-3) to (7,-0.375);
		\draw[dotted] (-7,-3) to (0,-3);
		\draw[dotted,blue] (7,-0.375) to (8,0);
		
		\draw[dotted] (-7,3) to (5.4444,0.66675) to (7,0.375);
		\draw[dotted] (-7,-3) to (5.4444,-0.66675) to (7,-0.375);
		\draw[blue] (7,0.375) to (9,0);
		\draw[blue] (7,-0.375) to (9,0);

		\coordinate (a) at (-1,-0.25); \coordinate (b) at (0,0); \coordinate (c) at (-1,0); 
		\pic[draw,"$\theta_{in}$", draw=black,<->,angle eccentricity=1.3,angle radius=1.5cm]{angle=c--b--a};
		
		\coordinate (d) at (8,1.75); \coordinate (e) at (7,1.75); \coordinate (f) at (9,4); 
		\pic[draw,"$\theta_{out}$", draw=black,<->,angle eccentricity=1.3,angle radius=1.5cm]{angle=d--e--f};	
		
		\coordinate (g) at (7,0); \coordinate (h) at (5.4444,0); \coordinate (i) at (7,1.75); 
		\pic[draw,"$\theta_{out}$", draw=black,<->,angle eccentricity=1.3,angle radius=0.5cm]{angle=g--h--i};	
		
		\draw[dotted,red] (5.4444,0) to (7,1.75);
		
		\draw[red] (5.4444,0.66675) circle (1pt);
		\draw[dotted,red] (5.4444,0.66675) to (7,2.125);
		\draw[dotted,red] (5.4444,-0.66675) to (7,1.375);
		
		\draw[dotted] (5.4444,-3) to (5.4444,3) ;   
		\draw[-|] (5.4444,3) to (5.4444,0.66675);% node[below,anchor=south,rotate=90]{Exit Pupil} ;  %pupilla uscita
		\draw (5.4444,0) node[below,anchor=south,rotate=90]{Exit Pupil} ;
		\draw[-|] (5.4444,-3) to (5.4444,-0.66675);

		\draw[dashed, <->] (-7,-3.5) -- (9,-3.5) node[above, midway]{$f_{eq}$};
		\draw[dashed, <->] (5.4444,-2.5) -- (9,-2.5) node[below, midway]{$d_p$};

		%%% LEGEND
		\draw[blue, thick] (-5.5,1.5) to (-4.5,1.5) node[right]{on axis source};
		\draw[red, thick] (-5.5,1) to (-4.5,1) node[right]{off axis source};
		
	\end{tikzpicture}
	\caption{Schematic diagram of a telescope with an optical instrument that increases the equivalent focal length, $f_{eq}$, and with the exit pupil closer to the focal plane than the entrance pupil ($d_p<f_{eq}$) so that the non-telecentric angle, $\theta_{out}$, is greater than the entrance radiation angle $\theta_{in}$}\label{figure:fovnta}
\end{figure}
We call $\theta_{in}$ the tilt of the incoming beam of an off-axis source; $\theta_{out}$ the non-telecentric angle that is the angle of the chief ray with respect to the optical axis after the last optical element; $d_p$ the distance of the exit pupil from the focal plane, $D_e$ the size of the entrance pupil, $f_{eq}$ the equivalent focal length and $F$ the focal ratio. \\
In Figure \ref{figure:fovnta} we show the non-telecentric angle $\theta_{out}$   from the diagram that, in small angle approximation, 
\begin{equation}
    d_p \theta_{out}=f_{eq}\theta_{in}
\end{equation}
Recalling the definition of F-number the equivalent focal length $f_{eq}$ is $f_{eq}=D_e F$ where $D_e$ is the size of the entrance pupil. In our scheme, the size of the entrance pupil equals the diameter of the first optical element. Thus 
%the non-telecentric angle $\theta_{out}$ is directly proportional to the entrance angle ($\theta_{in}$),  the size of entrance pupil ($D_e$) and to the focal ratio ($F$), instead it is inversely proportional to the distance of the exit pupil ($d_p$).
\begin{equation}
	\theta_{out}=\theta_{in} D_e \frac{F}{d_p}
\end{equation}

\section{NON TELECENTRIC ANGLES SPACE}\label{sec:ntaspace}
%%%%%%%%%%%%%%%%%%%%%%%%%%%%%%%%%%%%%%%%%%%%%%%%%%%%%%%%%%%%%%%%%%%%%%%%%%%%%%%%%%%%%%%%%%%%%%%%%%%%%%%%%%%%%%%%%%%%%%
%%%%%%%%%%%%%%%%%%%%%%%%%%%%%%%%%%%%%%%%%%%%%%%%%%%% NTA SPACE %%%%%%%%%%%%%%%%%%%%%%%%%%%%%%%%%%%%%%%%%%%%%%%%%%%%%%%
%%%%%%%%%%%%%%%%%%%%%%%%%%%%%%%%%%%%%%%%%%%%%%%%%%%%%%%%%%%%%%%%%%%%%%%%%%%%%%%%%%%%%%%%%%%%%%%%%%%%%%%%%%%%%%%%%%%%%%

In the Figure \ref{nonNormality} we show the non-normal space in which to represent the pupil on the filter \cite{Cremillo} in terms of angular measurements. \\
The angle $\theta_{out}$ defines the deviation from the normal of the chief ray of the converging cone of light relative to a particular source, while $\rho_{0}$ and $\epsilon_{\rho_{0}}$ represent the half-aperture of the converging beam and the obstruction, respectively. The $\rho_{0}$ is defined in Figure \ref{lightcone} and depends on $F$: if the focal number decreases the semi-aperture of the beam $\rho_0$ increases monotonically and asymptotically to $\pi /2$.\\ For the calculation of the contribution that each filter area element makes to the total flux reaching the sensor, we partitioned the pupil footprint with a fine grid of squares (exaggerated in figure \ref{nonNormality}) with a density of about 15000 samples over the entire surface area.
%\footnote{In reality, all calculations are performed on half of the footprint given the symmetry conditions}. 

\begin{figure}
\centering
\begin{tikzpicture}[scale=0.8]
	\draw 
	(0,0) coordinate (a) 
	-- (4,8) coordinate (b) node[above]{Focal point}
	-- (4,0) coordinate (c) node[midway,right]{$f$} node[below  left]{Exit Pupil}
	pic["$\rho_{0}$", draw=orange,<->,angle eccentricity=1.3,angle radius=3cm]{angle=a--b--c};
	\draw (8,0) -- (4,8);
	\draw (0,0) -- (8,0) ;
	\draw[<->] (4,-0.2) -- (8,-0.2) node[midway, below]{$\frac{D}{2}$};
	\draw (-2.5,7) node[right]{$\tan{\rho_{0}}=\frac{D/2}{f}$, $F=\frac{f}{D}$};
	\draw (-2.5,6) node[right]{$\Rightarrow \: \rho_{0}=\arctan{\frac{1}{2F}}$};
\end{tikzpicture}
	\caption{Converging beam} \label{lightcone}	
\end{figure}
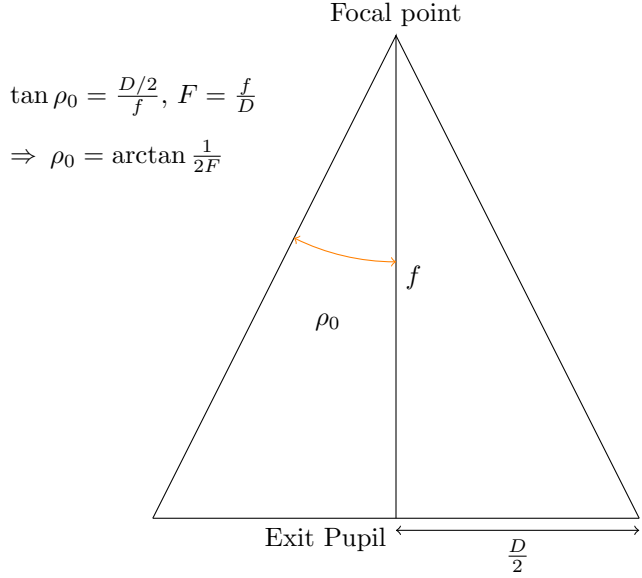

%\begin{figure}
%	\centering
%	\input{./TeX_files/AngleOnFilter.tex}
%	\caption{The radiation in the converging beam area affects the filter with a spectrum of angles from 0 to the center and increasing outward}%incident-light angles on the filter. The angle to the normal increases going outer} 
%	\label{AnglesOnFilter}	
%\end{figure}

\begin{figure}
 \centering
    \begin{subfigure}[]{0.48\textwidth}
    \centering
        \resizebox{\textwidth}{!}{
        \begin{tikzpicture}[>=stealth, inner sep=0pt, outer sep=2pt]
                    	%GRID
                	%\draw[step=1cm,gray!50,very thin,dashed] (-3,-5) grid (7,5);
                	\foreach \x in {-3,-2,...,7}{
		            \draw[dashed,black!40,very thin] (\x,-5)--(\x,5);
	            	\draw[dashed,black!40,very thin] (-3,\x-2)--(7,\x-2);
	                 }   

                	%AXIS
                	\draw[thick, ->] (-3.5,0) -- (8,0) node[below,font=\large]{\textbf{$\phi_{x}$}};
                	\draw[thick, ->] (0,-5.5) -- (0,6) node[left,font=\large]{\textbf{$\phi_{y}$}};
		
                	%PUPIL
                	\draw (2,0) circle (5);
                	\draw (2,0) circle (1.5);
	
                	%LITTLE SQUARE
                	\draw[draw=black!50,pattern=dots,pattern color=orange!50] (3,3)--(3,4)--(4,4)--(4,3)--(3,3);
                	\draw[<->] (2.9,3)--(2.9,4) node[midway,left,font=\large]{$d\phi_{y}$};
                	\draw[<->] (3,2.9)--(4,2.9) node[midway,below,font=\large]{$d\phi_{x}$};
                	%phi
                	\draw[very thick,->,green!50] (0,0)--(3.5,3.5) node[right,black,font=\boldmath]{\textbf{$\theta$}};
                	%nta
                	\draw[very thick,->,orange] (0,0) -- (2,0) node[below left,black,font=\boldmath]{\textbf{$\theta_{out}$}};
                	\draw[very thick,->,magenta] (2,0) -- (5,-4) node[below,black,font=\boldmath]{\textbf{$\rho_{0}$}};
                	\draw[very thick,->,purple] (2,0) -- (2,-1.5) node[below,black,font=\boldmath]{\textbf{$\epsilon_{0}$}};	
            \end{tikzpicture}
                	
        }
        \caption{The pupil on the non-normal space}   
        \label{nonNormality}
    \end{subfigure}
    \begin{subfigure}[]{0.48\textwidth}
        \centering
        \resizebox{\textwidth}{!}{
        \begin{tikzpicture}[>=stealth, inner sep=0pt, outer sep=2pt]
	%GRID
	%\draw[step=1cm,gray!50,very thin,dashed] (-3,-5) grid (7,5);
	\foreach \x in {-3,-2,...,7}{
		\draw[dashed,black!40,very thin] (\x,-5)--(\x,5);
		\draw[dashed,black!40,very thin] (-3,\x-2)--(7,\x-2);
	}
	%AXIS
	\draw[thick, ->] (-3.5,0) -- (8,0) node[below,font=\large]{\textbf{$\phi_{x}$}};
	\draw[thick, ->] (0,-5.5) -- (0,6) node[left,font=\large]{\textbf{$\phi_{y}$}};		
	%PUPIL
	\draw (2,0) circle (5);
	\draw (2,0) circle (1.5);
	%LITTLE SQUARES
	\draw[draw=black,pattern=dots,pattern color=cyan!50] (3,3)--(3,4)--(4,4)--(4,3)--(3,3);
	\draw[draw=black,pattern=dots,pattern color=orange!50] (-1,-1)--(-1,-2)--(-2,-2)--(-2,-1)--(-1,-1);
	%\draw[<->] (2.9,3)--(2.9,4) node[midway,left,font=\large]{$d\phi_{y}$};
	%\draw[<->] (3,2.9)--(4,2.9) node[midway,below,font=\large]{$d\phi_{x}$};
	%phi
	\draw[very thick,->,purple] (0,0)--(-1.5,-1.5) node[below,black,font=\boldmath]{\textbf{$\theta$}};
	\draw[very thick,->,cyan] (0,0)--(3.5,3.5) node[right,black,font=\boldmath]{\textbf{$\theta'$}};
	%\draw[] (4.2426,0) arc [start angle=0,end angle=100.3, radius=4.2426];
	%\draw[] (5.6568,0) arc [start angle=0,end angle=60.913, radius=5.6568];
	%\draw[] (4.2426,0) arc [start angle=0,end angle=-100.3, radius=4.2426];
	%\draw[] (5.6568,0) arc [start angle=0,end angle=-60.913, radius=5.6568];
	\clip (2,0) circle (5);
	\filldraw[pattern=dots,pattern color=cyan,even odd rule]   (0,0) circle (5.6568) (0,0) circle (4.2426);
	%REVERSECLIPPING
	\clip[insert path={(-99cm,-99cm) rectangle (99cm,99cm)}] (2,0) circle (1.5);
	\filldraw[pattern=dots,pattern color=orange,even odd rule]   (0,0) circle (1.4142) (0,0) circle (2.8284);
	%nta
	%\draw[very thick,->,orange] (0,0) -- (2,0) node[below left,black,font=\boldmath]{\textbf{$\theta_{0}$}};
	%\draw[very thick,->,magenta] (2,0) -- (5,-4) node[below,black,font=\boldmath]{\textbf{$\rho_{0}$}};
	%\draw[very thick,->,purple] (2,0) -- (2,-1.5) node[below,black,font=\boldmath]{\textbf{$\epsilon\rho_{0}$}};	
\end{tikzpicture} 
       }
        \caption[Isofiltering rings]{Isofiltering rings.} 
        \label{isofilter}
    \end{subfigure}
\caption{The entrance pupil as footprinted on the filter, represented on a non-normality space where $\theta_{out}$ represents the non-telecentric angle of a specific point in the field of view and $\rho_{0}$ represents the light-cone aperture as identified in figure \ref{lightcone}. In the panel on the right are highlighted two isofiltering rings, all the elements inside a ring present the same transmission curve.}
\label{fig:nonnormalspace}
\end{figure}
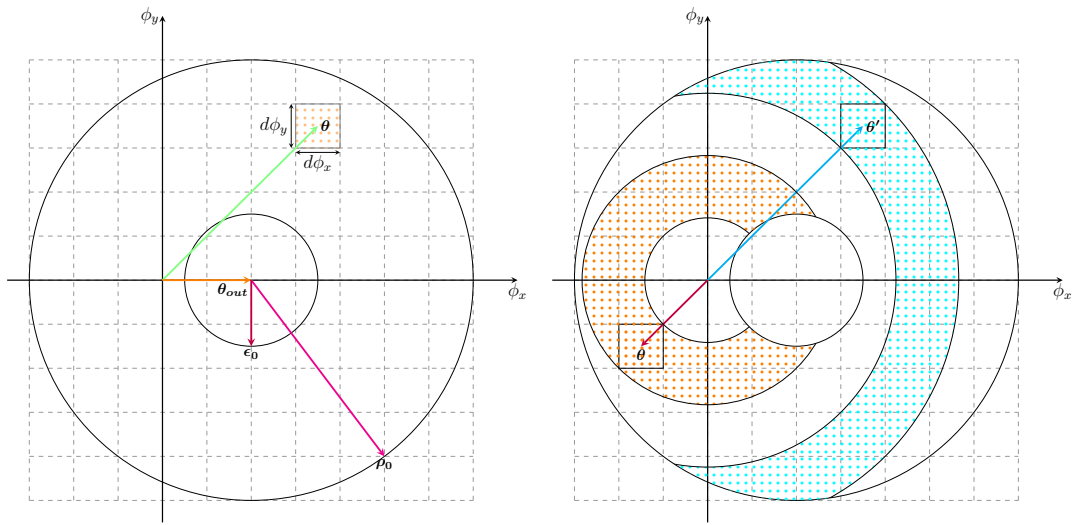

\begin{figure} 
	\centering
	\includegraphics[width=0.5\textwidth]{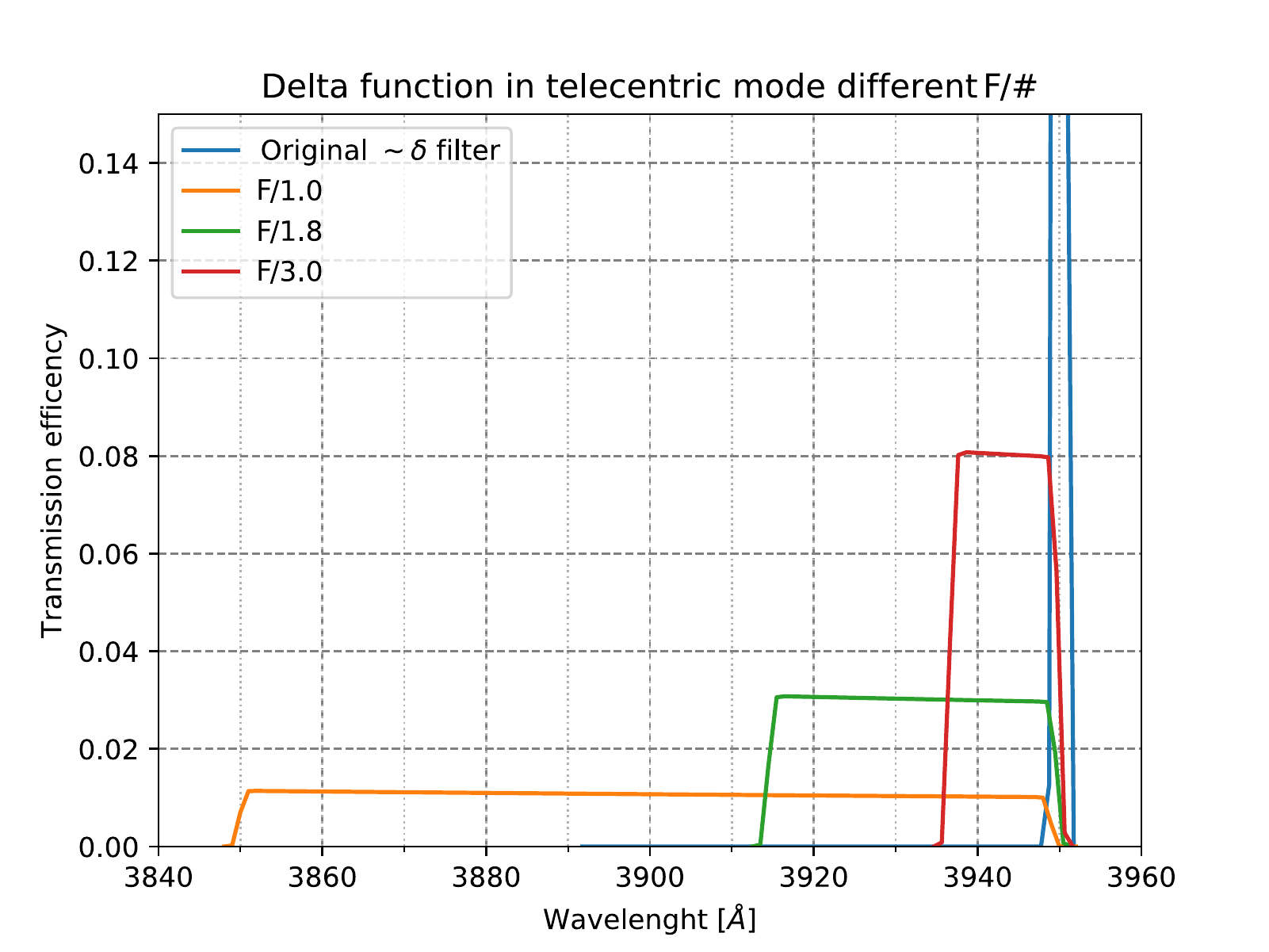}
	\caption{A very narrow nominal filter ($\sim\delta$) is dramatically deformed at different focal ratios because each filter element receives radiation at a different angle. It can be seen that the efficiency increases for shorter wavelengths, for the reasons explained in the text regarding isofiltering rings.}\label{deltafunction}
\end{figure}

Each element of area $d\phi_{x}d\phi_{y}$ (fig. \ref{nonNormality}) receives a flux equal to the overall intensity of the beam (which we consider normalized to 1) divided by the number of samples\footnote{Computationally, not knowing \textit{a priori} the exact number of samples, since beam aperture and pupil shape may vary, we considered each element with flux equal to 1, then provided for normalization of the integral.} Each sample represents a different filter than the nominal one since it receives radiation at a non-normal angle to the surface, which causes a backward shift in the passband as explained in the appendix \ref{appen:optical}.\\
Hence equation \ref{math:lambdatheta}
\begin{equation}\label{math:lambdatheta}
\lambda(\theta)=\lambda_0 \sqrt{1-\frac{\sin^2(\theta)}{\mu_*^2}}
\end{equation}
where $\lambda_0$ is the central wavelength of the nominal transmission curve for a parallel beam, $\mu_*$ is the effective refractive index, defined in Pidgeon \cite{Pidgeon:64} and shown in appendix \ref{sec:alldielectric} and $\theta$ is the previous defined $\theta_{out}$. This function represents the behavior of the central wavelength as the angle of incidence of the main beam changes for an interferential filter of the multiple thin-layer type.\\
In the non-normal space, there are isofiltering curves (circumferences in the telecentric case, otherwise arcs of circumference) that we can approximate with rings as in figure \ref{isofilter}. It is clear that in the telecentric case, outermost rings present a larger surface area than the innermost ones, going to result in a preponderance of the most translated wavelengths.\\
One can see the effect in Figure \ref{deltafunction} where we can see the collapse of the transmission efficiency for a very narrow filter that is \textit{dragged} over shorter wavelengths with a slow growth toward the minimum length\footnote{Here as in all cases where not otherwise specified the simulations are carried out considering a relative obstruction $\epsilon=0.1$. Larger obstructions reduce the angle spread}. \\
 %If we assume that each infinitesimal filter element retains the same shape of the nominal transmission curve, simply with a central wavelength shifted according to the \ref{math:lambdatheta}, the integrated flux over the entire surface is preserved.\ This result is well established in the literature as long as the design of the filter thin layers has taken into account the possibility that the surface is reached by non-normal radiation while preserving the shape of the transmission curve \parencite[364-367]{macleod2010thin}. \\
 
\begin{figure}
	\centering
	\includegraphics[width=0.9\textwidth]{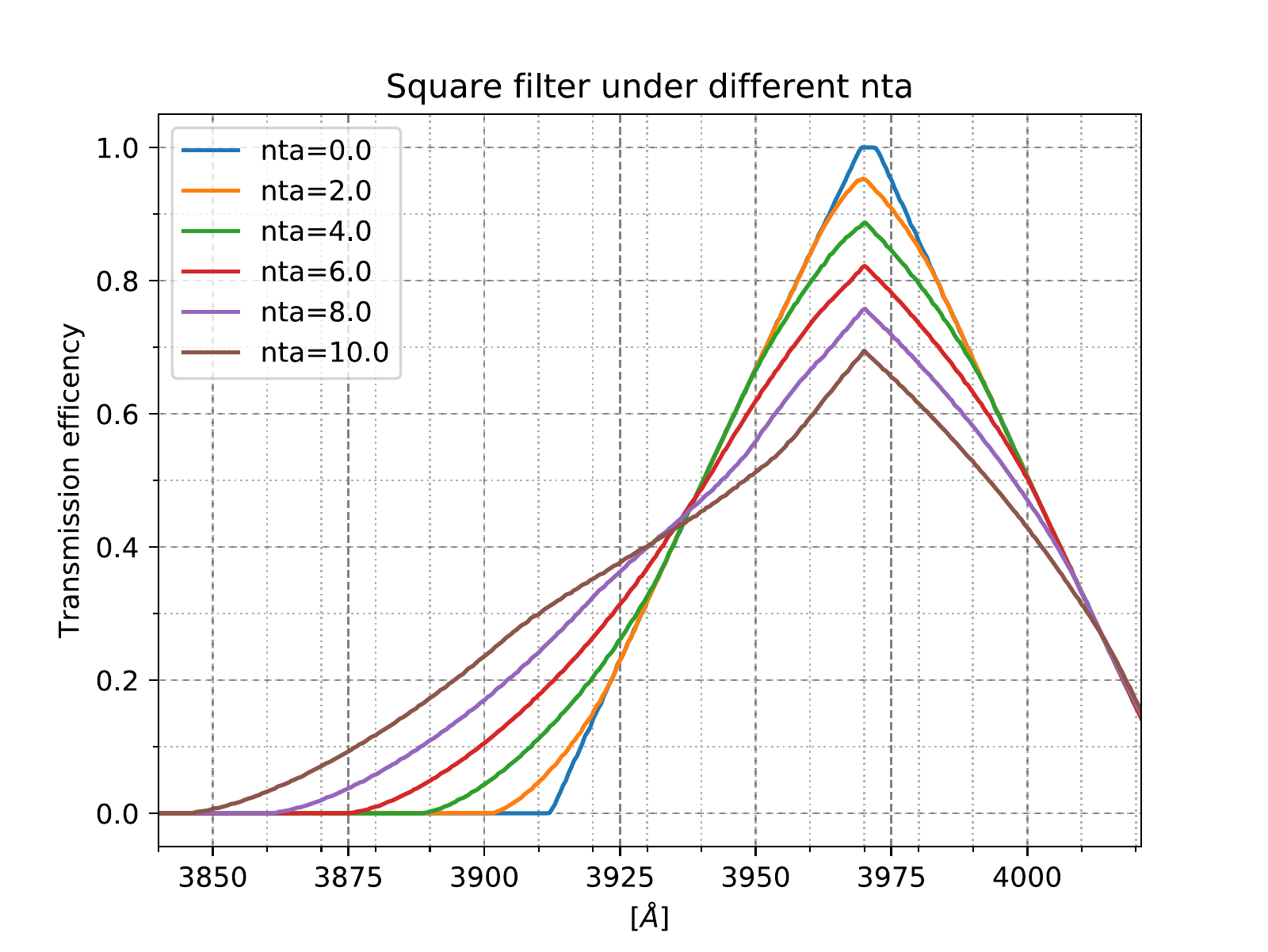}
	\caption{Deformation of a square filter of width 60\AA\ at different non-telecentric angles with F/1.4. As appreciable in the table \ref{table:squarefilternta}}\label{figure:squarefilternta}
\end{figure}

 The integral over the entire filter surface is therefore the result of spreading the contribution of individual area elements over a wavelength range. The wider the wavelength range over which the contribution of the individual area elements is spread, that is, the wider the aperture of the light cone (the smaller the F-number), the more the transmission efficiency for a given wavelength decreases. Each concentric circle thus represents a different filter: in the case depicted in Figure \ref{deltafunction} adjacent rings have transmission curves that hardly overlap because of the particularly narrow band. \\
In the non-telecentric case, when $\theta_{out}$ is larger than the angular size of the obstruction $\epsilon_0$, the filter is reached by radiation normal to the surface. Whereas in the telecentric case it is excluded from it because it falls into the obstruction. This leads to deformations of the integrated filter that are less linear\footnote{for example in fig. \ref{isofilter} the orange and cyan areas are comparable.}, as illustrated in Figure \ref{figure:asym-nta} and Figure \ref{figure:squarefilternta}. 

\begin{figure}
	\subfloat[]{\includegraphics[width=0.5\textwidth]{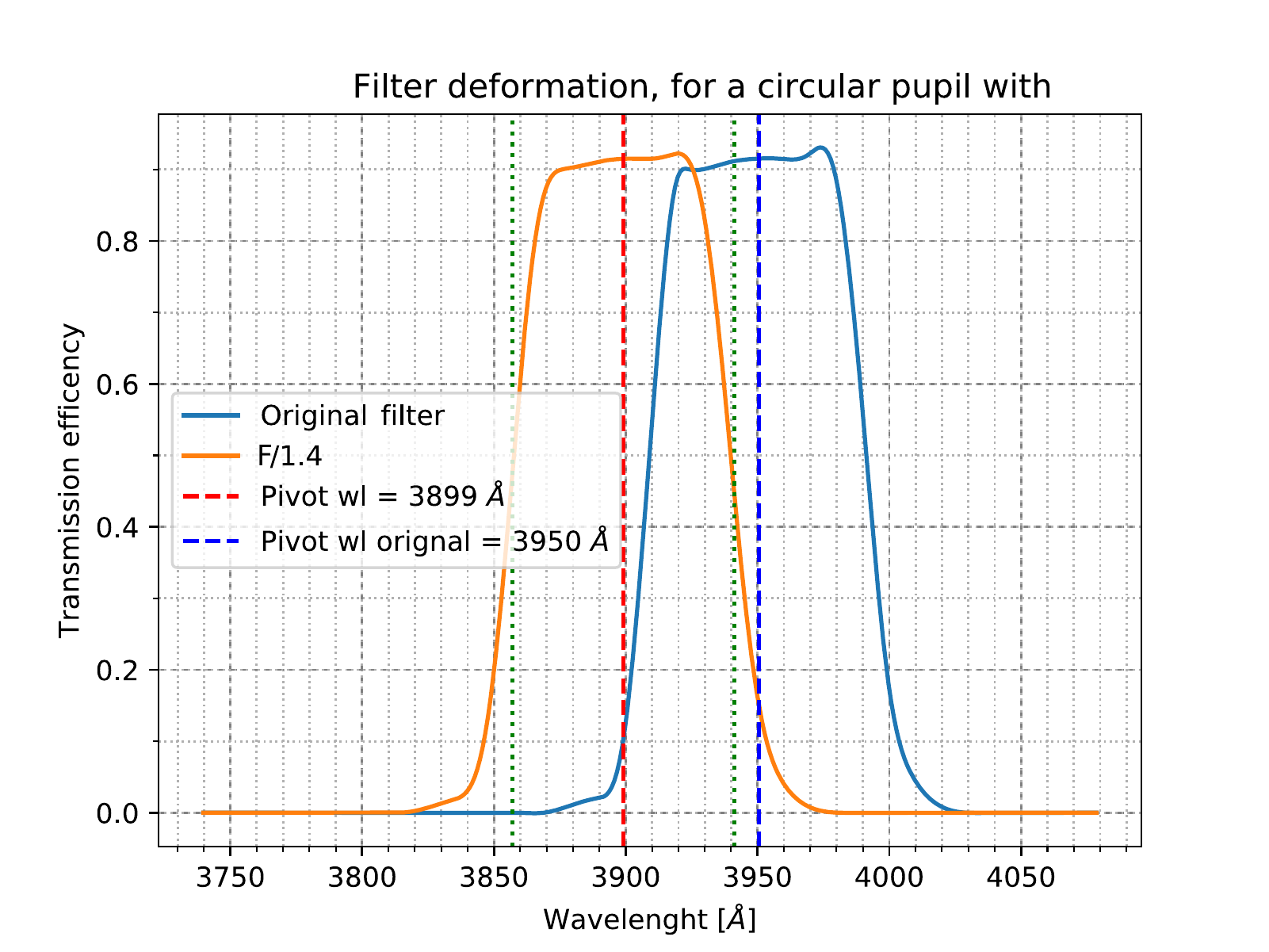}}
	\subfloat[]{\includegraphics[width=0.5\textwidth]{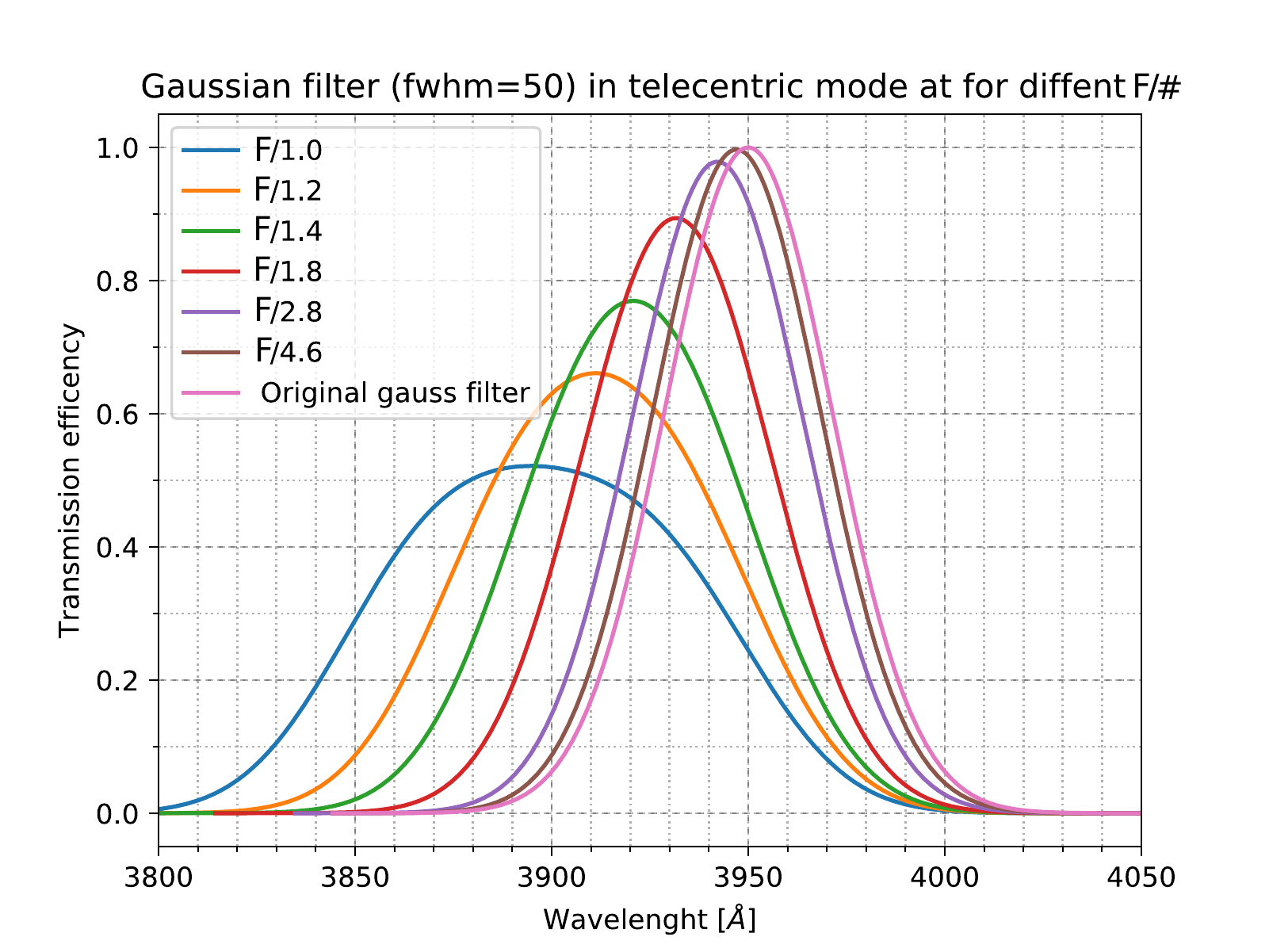}}
	\caption{Filter transmission's shape is deformed (shifted and smoothed) under the effect of a non-constant incidence angle. The extension of the aperture angles is related to the focal ratio.}
	\label{deformed}
\end{figure}

Interferential filters then respond to different F-numbers as depicted in figure \ref{deformed} (b). The graph was constructed from a Gaussian filter with a $FWHM=50$\AA: the filter widens and the transmission efficiency for each wavelength decreases as the focal ratio decreases. \\
A first qualitative look, through the figures \ref{deformed}(b) and \ref{deltafunction}, already gives us some indications. For each F-number there is a minimum value of the FWHM that can be reached however narrow the filter, for $F/1$ we are talking about almost $100$\AA. Similarly, however, we can see that for sufficiently large F-numbers the strain effect is small and very dependent on the initial shape of the filter.

\begin{figure}
	\subfloat[]{\includegraphics[width=0.5\textwidth]{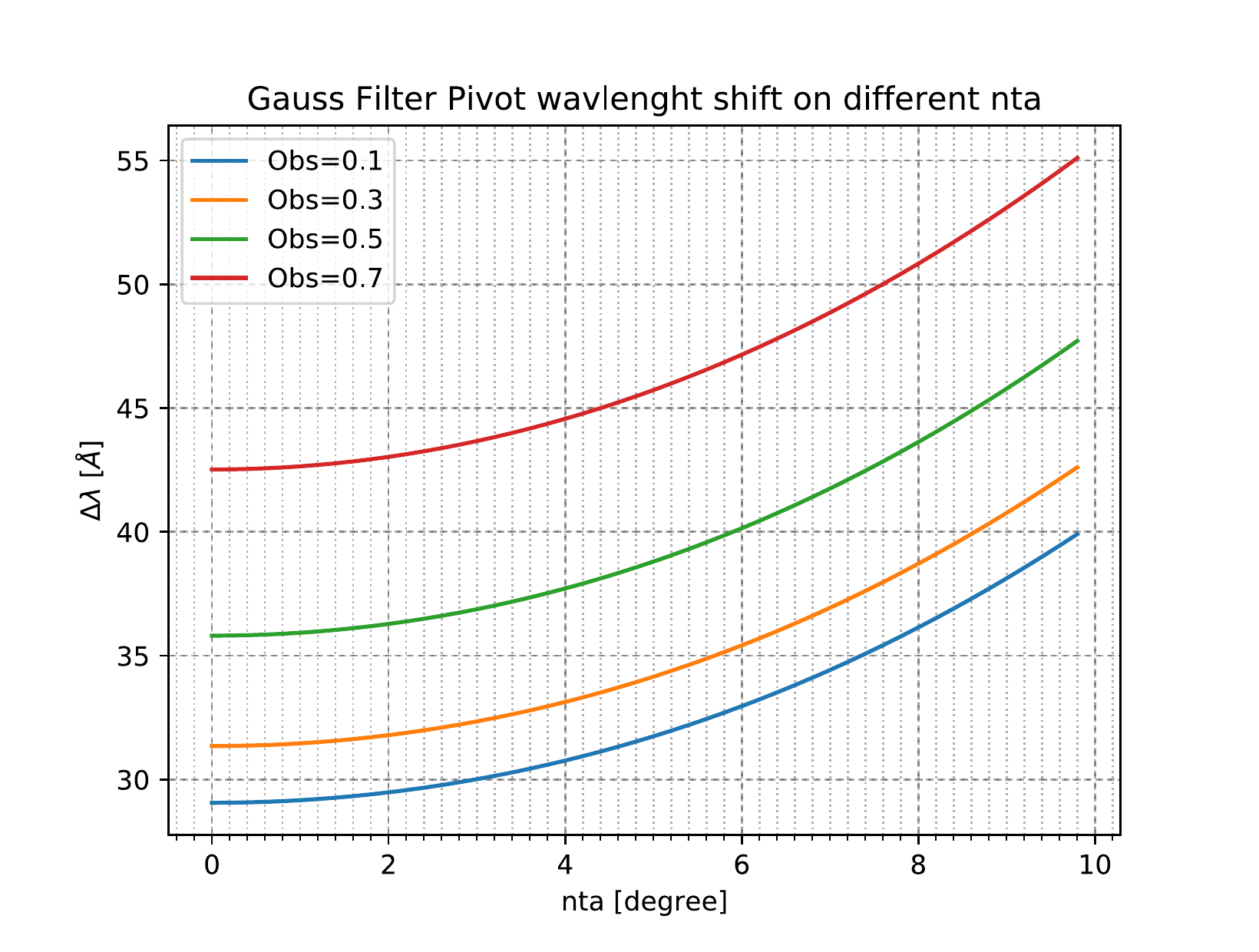}}
	\subfloat[]{\includegraphics[width=0.45\textwidth]{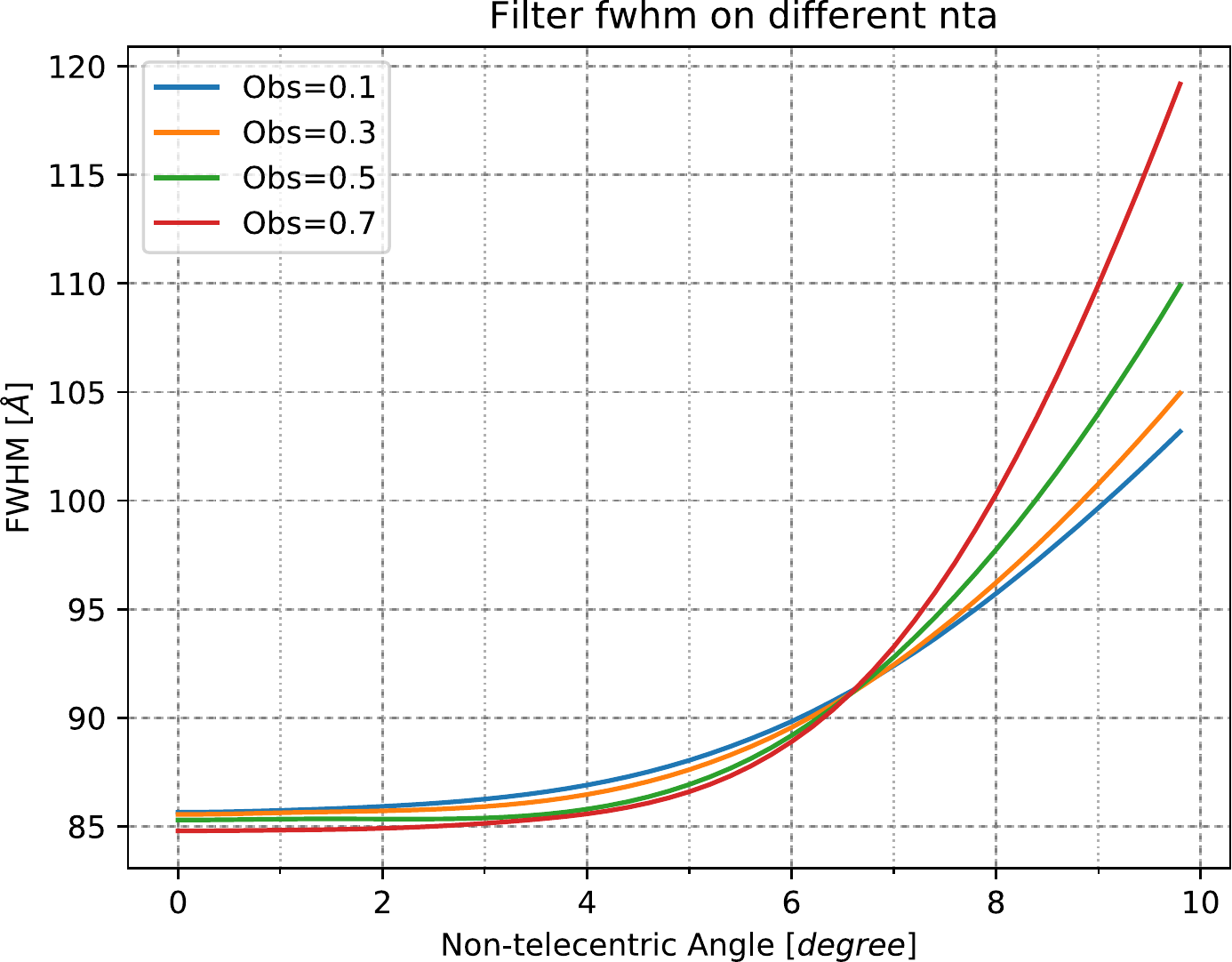}}
	\caption{The integrated transmission curve central wavelength varies with non-telecentric angle (panel a). A variation of position in the field of view corresponds to a variation of the non-telecentric angle. Both pivotal wavelength and FWHM (panel b) vary, possibly dramatically.} 
	\label{nta}
\end{figure}

\begin{figure}
	\centering
	\includegraphics[width=0.7\textwidth]{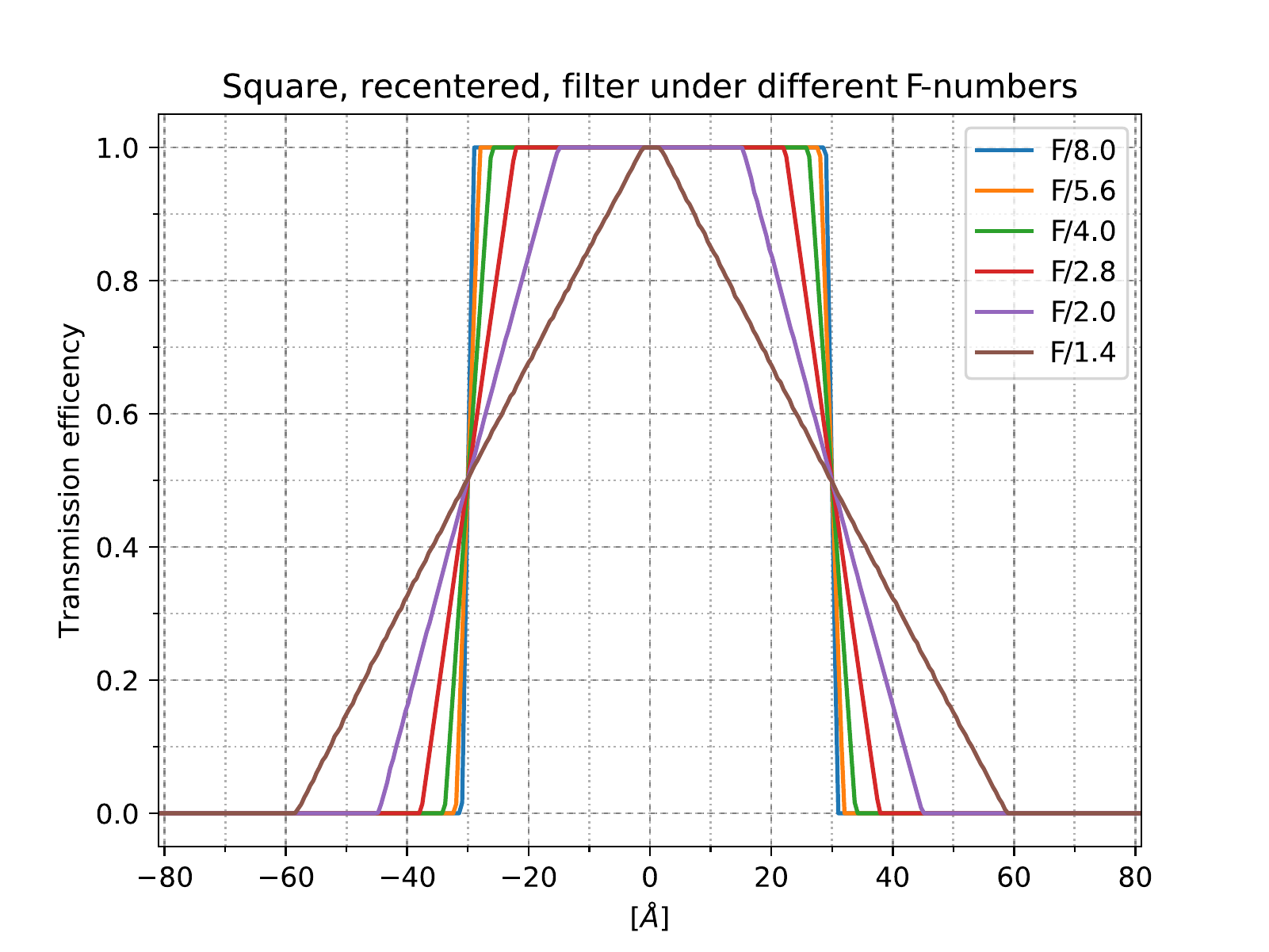}
	\caption{Deformation of a square filter of width 60 \AA, re-centered at 0\AA, from which we see the conservation of the FWHM and simultaneously the decrease of the flux falling within the FWHM. As appreciable in the table \ref{table:squarefilter}}\label{figure:squarefilter}
\end{figure}

\section{F-NUMBER RESPONSE}\label{sec:fnumeberres}
%%%%%%%%%%%%%%%%%%%%%%%%%%%%%%%%%%%%%%%%%%%%%%%%%%%%%%%%%%%%%%%%%%%%%%%%%%%%%%%%%%%%%%%%%%%%%%%%%%%%%%%%%%%%%%%%%%%%%%
%%%%%%%%%%%%%%%%%%%%%%%%%%%%%%%%%%%%%%%%%%%%%%% F# RESPONSE %%%%%%%%%%%%%%%%%%%%%%%%%%%%%%%%%%%%%%%%%%%%%%%%%%%%%%%%%%
%%%%%%%%%%%%%%%%%%%%%%%%%%%%%%%%%%%%%%%%%%%%%%%%%%%%%%%%%%%%%%%%%%%%%%%%%%%%%%%%%%%%%%%%%%%%%%%%%%%%%%%%%%%%%%%%%%%%%%
For a more systematic study, we constructed a set of filters with a square transmission curve\footnote{zero transmission everywhere and transmission one in the passband, with a very rapid variation} and passbands of 10, 20, 40, 60, 80, and 100\AA of which to study the deformation from the qualitative point of view and especially the quantitative behavior of some parameters: position of the pivot wavelength\footnote{the length that divides the total flux in half.}, the FWHM, the transmission peak and what we have called $I_{ratio}$, the ratio of the flux that falls within the FWHM to the total flux expressed as a percentage, i.e., an indicator that in a complementary way highlight the relevance of the tails in the shape of the filter under analysis.\\
In figure \ref{figure:squarefilter} one can see the strain effect on the square filter of width $60$\AA originally centered at $4000$\AA for different focal ratios. In the image, each shape of the integrated transmission curve has been shifted to $\lambda_{Piv}=0$ to highlight the conservation of the FWHM with different focal ratios and the simultaneous decrease in the flux share that can be found within the FWHM itself, a phenomenon that justifies the need to introduce the parameter $I_{ratio}$. \\
Instead, in table \ref{table:squarefilter} we find the quantitative treatment extended to the whole set of defined filters. We have reported the data for a filter set centered on $4000$\AA on the left and on $2000$\AA on the right, the first column represents the percent retreat of the pivot wavelength from the initial 4000 or 2000\AA position, the second column reports the relative change in FWHM from the nominal filter set, the third the previously defined $I_{ratio}$ ratio, and the last column reports the maximum peak transmission. \\
From the analysis we can retrieve:
\begin{itemize}
    \item the relative blueshift for each F-number is characteristic of the F-number itself and the same for each filter: less than $0.05\%$ for F/8 and greater, up to $0.7\%$ for $F/1.4$, which in the case centered on $4000$\AA results in absolute backwardness of $28$\AA;
    \item the FWHM is always conserved ( $\frac{\Delta fwhm}{FWHM}=0$) for sufficiently large (for F/1.4 it starts to occur between $40$\AA and $60$\AA) nominal filters, as in the case from figure \ref{figure:squarefilter};
    \item the preservation of the FWHM does not correspond to the equivalence of the deformed filter with the nominal filter. We witness a strong degradation of the characteristics at the edges of the transmission curve, which in the case of figure \ref{figure:squarefilter} is transformed from square to triangular. The tails outside the FWHM then become significant allowing the transmission of radiation well beyond the limit of the nominal filter. 
\end{itemize}

We are dealing with transformations of the transmission curve that are manageable in the design phase: in the worst-case scenario present in the table \ref{table:squarefilter}, $FWHM=60$\AA and $F/1.4$, we still continue to have almost $80$\% of the flux inside the FWHM. Of course, we must remember that, for each focal ratio and proportional to the pivot wavelength, we have a minimum FWHM below which we cannot go: it will therefore be unrealistic to try to construct an interferential filter below $57$\AA ($5. 7x10$\AA tab. \ref{table:squarefilter}) for instruments with F/1.4 with $\lambda_0=4000$\AA (or $28$\AA with $\lambda_0=2000$\AA, or even by $113$\AA for $\lambda_0=8000$\AA) or below $30$\AA for F/2\footnote{we are always operating with an obstruction of 0.1}.

\begin{table}\centering
	\begin{tabular}{l|l}
		\begin{tabular}{lp{1.2cm}|lp{1.2cm}ll}
			\multicolumn{2}{c}{obs=0.1} & \multicolumn{4}{c}{4000\AA}\\ 
			&F/ & $\frac{\Delta\lambda_p}{\lambda_0}\%$ & $\frac{\Delta FWHM}{FWHM}$ & Iratio & $max_{tr}$ \\ \hline\hline
			\multirow{6}{*}{\rotatebox[origin=c]{90}{10 \AA}}
			&8.0 & 0.0 & 0.00 & 95.01 & 1.00 \\ 
			&5.6 & 0.1 & 0.00 & 90.02 & 1.00 \\ 
			&4.0 & 0.1 & 0.00 & 80.63 & 1.00 \\ 
			&2.8 & 0.2 & 0.55 & 83.63 & 0.65 \\ 
			&2.0 & 0.4 & 1.94 & 91.09 & 0.35 \\ 
			&1.4 & 0.7 & 4.67 & 95.19 & 0.18 \\ \hline %56.7\AA
			\multirow{6}{*}{\rotatebox[origin=c]{90}{20 \AA}}
			&8.0 & 0.0 & 0.00 & 97.50 & 1.00 \\ 
			&5.6 & 0.1 & 0.00 & 95.01 & 1.00 \\ 
			&4.0 & 0.1 & 0.00 & 90.31 & 1.00 \\ 
			&2.8 & 0.2 & 0.00 & 80.56 & 1.00 \\ 
			&2.0 & 0.4 & 0.47 & 82.72 & 0.68 \\ 
			&1.4 & 0.7 & 1.83 & 90.77 & 0.36 \\ \hline%56.6\AA
			\multirow{6}{*}{\rotatebox[origin=c]{90}{40 \AA}}
			&8.0 & 0.0 & 0.00 & 98.75 & 1.00 \\ 
			&5.6 & 0.1 & 0.00 & 97.50 & 1.00 \\ 
			&4.0 & 0.1 & 0.00 & 95.16 & 1.00 \\ 
			&2.8 & 0.2 & 0.00 & 90.28 & 1.00 \\ 
			&2.0 & 0.4 & 0.00 & 81.50 & 1.00 \\ 
			&1.4 & 0.7 & 0.42 & 82.05 & 0.71 \\ \hline %56.8
			\multirow{6}{*}{\rotatebox[origin=c]{90}{60 \AA}}
			&8.0 & 0.0 & 0.00 & 99.17 & 1.00 \\ 
			&5.6 & 0.1 & 0.00 & 98.34 & 1.00 \\ 
			&4.0 & 0.1 & 0.00 & 96.77 & 1.00 \\ 
			&2.8 & 0.2 & 0.00 & 93.52 & 1.00 \\ 
			&2.0 & 0.4 & 0.00 & 87.67 & 1.00 \\ 
			&1.4 & 0.7 & 0.00 & 76.20 & 1.00 \\ \hline
			\multirow{6}{*}{\rotatebox[origin=c]{90}{80 \AA}}
			&8.0 & 0.0 & 0.00 & 99.38 & 1.00 \\ 
			&5.6 & 0.1 & 0.00 & 98.75 & 1.00 \\ 
			&4.0 & 0.1 & 0.00 & 97.58 & 1.00 \\ 
			&2.8 & 0.2 & 0.00 & 95.14 & 1.00 \\ 
			&2.0 & 0.4 & 0.00 & 90.75 & 1.00 \\ 
			&1.4 & 0.7 & 0.00 & 82.15 & 1.00 \\ \hline
			\multirow{6}{*}{\rotatebox[origin=c]{90}{100 \AA}}
			&8.0 & 0.0 & 0.00 & 99.50 & 1.00 \\ 
			&5.6 & 0.1 & 0.00 & 99.00 & 1.00 \\ 
			&4.0 & 0.1 & 0.00 & 98.06 & 1.00 \\ 
			&2.8 & 0.2 & 0.00 & 96.11 & 1.00 \\ 
			&2.0 & 0.4 & 0.00 & 92.60 & 1.00 \\ 
			&1.4 & 0.7 & 0.00 & 85.72 & 1.00 \\
			\hline
		\end{tabular}
		&
		\begin{tabular}{llll}
     \multicolumn{4}{c}{2000\AA}\\
		 $\frac{\Delta\lambda_p}{\lambda_0}\%$ & $\frac{\Delta FWHM}{FWHM}$ & Iratio & $max_{tr}$ \\ \hline \hline
		 0.0 & 0.00 & 97.44 & 1.00 \\ 
		 0.1 & 0.00 & 94.96 & 1.00 \\ 
		 0.1 & 0.00 & 90.24 & 1.00 \\ 
		 0.2 & 0.00 & 80.60 & 1.00 \\ 
		 0.4 & 0.47 & 82.77 & 0.68 \\ 
		 0.7 & 1.83 & 90.79 & 0.36 \\ \hline %(28.3)
		 0.0 & 0.00 & 98.72 & 1.00 \\ 
		 0.1 & 0.00 & 97.48 & 1.00 \\ 
		 0.1 & 0.00 & 95.12 & 1.00 \\ 
		 0.2 & 0.00 & 90.30 & 1.00 \\ 
		 0.4 & 0.00 & 81.51 & 1.00 \\ 
		 0.7 & 0.42 & 82.07 & 0.71 \\ \hline %(28.4)
		 0.0 & 0.00 & 99.36 & 1.00 \\ 
		 0.1 & 0.00 & 98.74 & 1.00 \\
		 0.1 & 0.00 & 97.56 & 1.00 \\
		 0.2 & 0.00 & 95.15 & 1.00 \\ 
		 0.4 & 0.00 & 90.75 & 1.00 \\ 
		 0.7 & 0.00 & 82.14 & 1.00 \\ \hline
		 0.0 & 0.00 & 99.57 & 1.00 \\ 
		 0.1 & 0.00 & 99.16 & 1.00 \\ 
		 0.1 & 0.00 & 98.37 & 1.00 \\ 
		 0.2 & 0.00 & 96.76 & 1.00 \\ 
		 0.4 & 0.00 & 93.83 & 1.00 \\ 
		 0.7 & 0.00 & 88.09 & 1.00 \\ \hline
		 0.0 & 0.00 & 99.68 & 1.00 \\ 
		 0.1 & 0.00 & 99.37 & 1.00 \\ 
		 0.1 & 0.00 & 98.78 & 1.00 \\ 
		 0.2 & 0.00 & 97.57 & 1.00 \\ 
		 0.4 & 0.00 & 95.38 & 1.00 \\ 
		 0.7 & 0.00 & 91.07 & 1.00 \\ \hline
		 0.0 & 0.00 & 99.74 & 1.00 \\ 
		 0.1 & 0.00 & 99.50 & 1.00 \\ 
		 0.1 & 0.00 & 99.02 & 1.00 \\ 
		 0.2 & 0.00 & 98.06 & 1.00 \\ 
		 0.4 & 0.00 & 96.30 & 1.00 \\ 
		 0.7 & 0.00 & 92.85 & 1.00 \\ \hline
		\end{tabular}
	\end{tabular}
	\caption{A set of filters centered on $4000$\AA\ on the left and on $2000$\AA\ on the right are deformed in different F-numbers} \label{table:squarefilter}
\end{table}

\section{Non telecentric angles response}\label{sec:ntaresponse}
\begin{figure}
	\centering
	\includegraphics[width=0.7\textwidth]{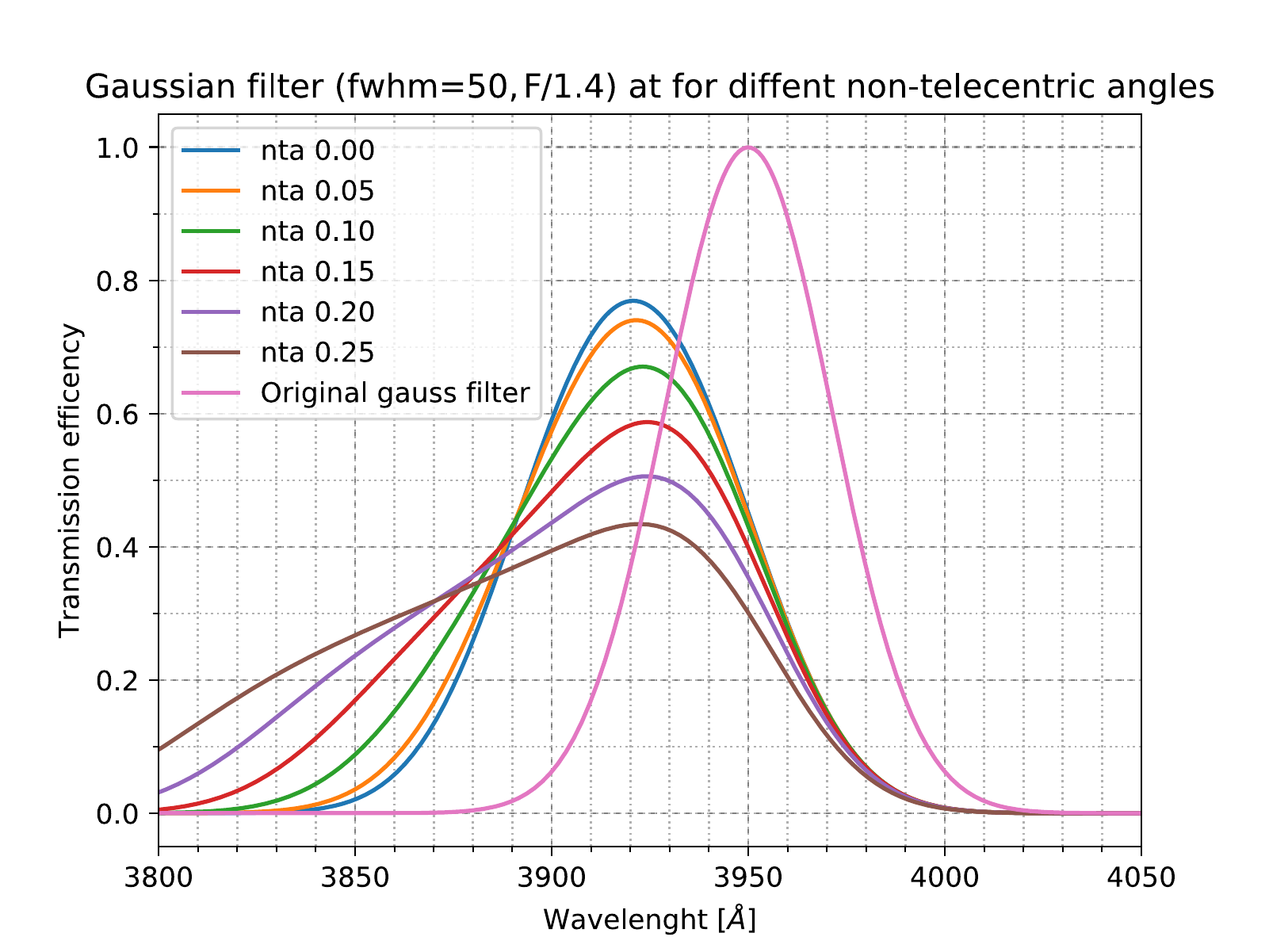}	
	\caption{Deformation of a Gaussian filter with different non-telecentric angles}\label{figure:asym-nta}
\end{figure}

\begin{figure}
	\centering
	\includegraphics[width=0.7\textwidth]{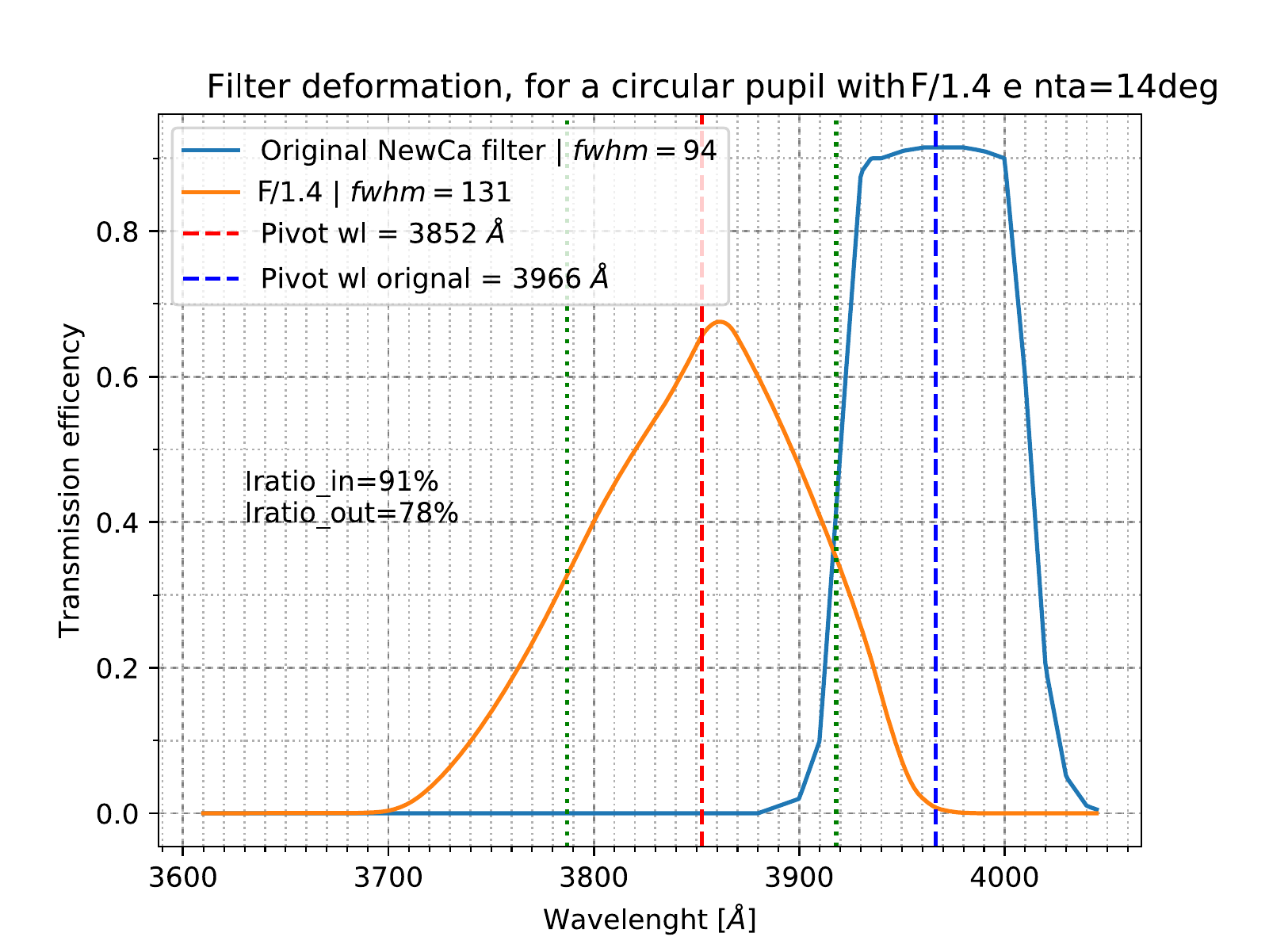}
	\caption{Filter deformation. Pivot length and $I_{ratio}$ are given.}\label{figure:Iratio}
\end{figure}

%%%%%%%%%%%%%%%%%%%%%%%%%%%%%%%%%%%%%%%%%%%%%%%%%%%%%%%%%%%%%%%%%%%%%%%%%%%%%%%%%%%%%%%%%%%%%%%%%%%%%%%%%%%%%%%%%%%%%%%
%%%%%%%%%%%%%%%%%%%%%%%%%%%%%%%%%%%%%%%%%%%%%%%  NTA RESPONSE %%%%%%%%%%%%%%%%%%%%%%%%%%%%%%%%%%%%%%%%%%%%%%%%%%%%%%%%%
%%%%%%%%%%%%%%%%%%%%%%%%%%%%%%%%%%%%%%%%%%%%%%%%%%%%%%%%%%%%%%%%%%%%%%%%%%%%%%%%%%%%%%%%%%%%%%%%%%%%%%%%%%%%%%%%%%%%%%%
As already briefly summarized in the previous section and illustrated in figure \ref{isofilter}, if the chief ray of the converging beam reaches the surface of the filter deviating from the normal the contribution of the individual surface elements gives the integrated transmission curve a deformation, relative to the nominal shape, that depends on the distance from the normal. In figure \ref{nta}(a) one can follow the deviation of the peak wavelength from the nominal wavelength of the filter in a Gaussian case, and in figure \ref{nta}(b), one can instead see how the $FWHM$ varies for a filter that has a nominal width of $85$\AA not very dissimilar to the $Ca_{new}$ filter used by J.-W. Lee \cite{Lee_2017}. 
The asymmetric deformation can be qualitatively appreciated in figure \ref{figure:asym-nta}. \\
Since the non-telecentric angle is proportional to the angular distance from the center of the field of view, the introduced deformation generates differential \textit{aberration}: sources at different positions within the FoV are differentially filtered. Although the transmission peak remains fairly stable, as visible in figure \ref{figure:asym-nta}, the center of gravity of the transmission curve shifts to shorter wavelengths and with a higher FWHM. To account for this deformation, we used the previously defined quantities: the pivot wavelength and $I_{ratio}$. In figure \ref{figure:Iratio} you can see the effect of deformation on a filter with the quantities just remembered indicated.

In table \ref{table:squarefilternta} we have the same parameters given in table \ref{table:squarefilter}, no longer varying the $F-number$ but rather the non-telecentric angle ($0\deg$ to $10\deg$) on an instrument with relative obstruction of 0.1. In addition, two cases are presented: on the left with F/1.4 and on the right with F/4. \\
The presence of a non-telecentric angle (we will denote it as \textit{nta} from now on) leads to differential aberration within the field of view, for any focal ratio and any FWHM, for two orders of reasons: 
\begin{itemize}
	\item there is always a greater blueshift of the pivot wavelength for larger \textit{nta} (thus toward the edge of the FoV)
	\item can originate for the $I_{ratio}$ a difference greater than 10\%, even for wide bandwidths and relaxed F-number as in the case of the $80$\AA-width filter with $F/4$ (last four columns of the fifth group of table rows \ref{table:squarefilternta}).
\end{itemize}
  The case from $60$\AA, visible in figure \ref{figure:asym-nta} and reported in the table, gives us an account of the asymmetry of the deformation introduced by the non-telecentric angle. Thus we can deduce the need for a detailed study of the instrument/scientific case coupling to characterize the most suitable filter.

\begin{table}
	\begin{tabular}{l|l}
		\begin{tabular}{ll|llll}
			&nta & $\frac{\Delta\lambda_p}{\lambda_0}\%$ & $\frac{\Delta FWHM}{FWHM}$ & Iratio & $max_{tr}$ \\ \hline\hline
			\multirow{6}{*}{\rotatebox[origin=c]{90}{10 \AA}}
			&0.0 & 0.7 & 4.67 & 95.19 & 0.18 \\ 
			&2.0 & 0.7 & 4.67 & 90.63 & 0.18 \\ 
			&4.0 & 0.7 & 4.61 & 83.57 & 0.18 \\ 
			&6.0 & 0.8 & 4.44 & 75.50 & 0.17 \\ 
			&8.0 & 0.8 & 4.28 & 67.85 & 0.17 \\ 
			&10.0 & 0.9 & 3.69 & 56.81 & 0.17 \\ \hline
			\multirow{6}{*}{\rotatebox[origin=c]{90}{20 \AA}}
			&0.0 & 0.7 & 1.83 & 90.77 & 0.36 \\ 
			&2.0 & 0.7 & 1.85 & 88.26 & 0.35 \\ 
			&4.0 & 0.7 & 1.82 & 81.53 & 0.35 \\ 
			&6.0 & 0.8 & 1.75 & 73.87 & 0.34 \\ 
			&8.0 & 0.8 & 1.72 & 66.92 & 0.33 \\ 
			&10.0 & 0.9 & 1.79 & 61.87 & 0.31 \\ \hline
			\multirow{6}{*}{\rotatebox[origin=c]{90}{40 \AA}}
			&0.0 & 0.7 & 0.42 & 82.05 & 0.71 \\ 
			&2.0 & 0.7 & 0.43 & 81.11 & 0.70 \\ 
			&4.0 & 0.8 & 0.46 & 77.85 & 0.67 \\ 
			&6.0 & 0.8 & 0.54 & 74.28 & 0.63 \\ 
			&8.0 & 0.8 & 0.67 & 72.07 & 0.58 \\ 
			&10.0 & 0.9 & 0.87 & 70.98 & 0.53 \\ \hline
			\multirow{6}{*}{\rotatebox[origin=c]{90}{60 \AA, fig.\ref{figure:squarefilternta}}}
			&0.0 & 0.7 & 0.00 & 76.20 & 1.00 \\ 
			&2.0 & 0.7 & 0.05 & 77.42 & 0.95 \\ 
			&4.0 & 0.8 & 0.11 & 77.34 & 0.89 \\ 
			&6.0 & 0.8 & 0.20 & 76.45 & 0.82 \\ 
			&8.0 & 0.9 & 0.33 & 75.94 & 0.76 \\ 
			&10.0 & 1.0 & 0.47 & 74.98 & 0.69 \\ \hline
			\multirow{6}{*}{\rotatebox[origin=c]{90}{80 \AA}}
			&0.0 & 0.7 & 0.00 & 82.15 & 1.00 \\ 
			&2.0 & 0.8 & 0.00 & 81.40 & 1.00 \\ 
			&4.0 & 0.8 & 0.00 & 79.32 & 1.00 \\ 
			&6.0 & 0.8 & 0.04 & 77.54 & 0.95 \\ 
			&8.0 & 0.9 & 0.12 & 76.65 & 0.89 \\ 
			&10.0 & 1.0 & 0.23 & 76.20 & 0.83 \\ \hline
			\multirow{6}{*}{\rotatebox[origin=c]{90}{100 \AA}}
			&0.0 & 0.7 & 0.00 & 85.72 & 1.00 \\ 
			&2.0 & 0.8 & 0.00 & 85.12 & 1.00 \\ 
			&4.0 & 0.8 & 0.00 & 83.37 & 1.00 \\ 
			&6.0 & 0.9 & 0.00 & 80.47 & 1.00 \\ 
			&8.0 & 1.0 & 0.02 & 77.57 & 0.98 \\ 
			&10.0 & 1.1 & 0.07 & 76.29 & 0.93 \\ \hline \hline
		\end{tabular}&
		\begin{tabular}{llll}
 $\frac{\Delta\lambda_p}{\lambda_0}\%$ & $\frac{\Delta FWHM}{FWHM}$ & Iratio & $max_{tr}$ \\ \hline \hline
 0.1 & 0.00 & 80.63 & 1.00 \\ 
 0.1 & 0.07 & 77.61 & 0.92 \\
 0.1 & 0.37 & 75.64 & 0.74 \\ 
 0.2 & 0.85 & 76.80 & 0.57 \\ 
 0.3 & 1.59 & 82.75 & 0.42 \\ 
 0.4 & 2.37 & 86.57 & 0.32 \\ \hline
 0.1 & 0.00 & 90.31 & 1.00 \\ 
 0.1 & 0.00 & 87.21 & 1.00 \\ 
 0.2 & 0.00 & 79.63 & 1.00 \\ 
 0.2 & 0.10 & 76.01 & 0.91 \\ 
 0.3 & 0.38 & 78.64 & 0.74 \\ 
 0.5 & 0.71 & 81.88 & 0.60 \\ \hline
 0.1 & 0.00 & 95.16 & 1.00 \\ 
 0.1 & 0.00 & 93.60 & 1.00 \\ 
 0.2 & 0.00 & 89.81 & 1.00 \\ 
 0.2 & 0.00 & 85.52 & 1.00 \\ 
 0.3 & 0.00 & 81.16 & 1.00 \\ 
 0.5 & 0.01 & 77.50 & 0.98 \\ \hline
 0.1 & 0.00 & 96.77 & 1.00 \\ 
 0.1 & 0.00 & 95.74 & 1.00 \\ 
 0.2 & 0.00 & 93.21 & 1.00 \\ 
 0.2 & 0.00 & 90.35 & 1.00 \\ 
 0.3 & 0.00 & 87.44 & 1.00 \\ 
 0.5 & 0.00 & 84.53 & 1.00 \\ \hline
 0.1 & 0.00 & 97.58 & 1.00 \\ 
 0.1 & 0.00 & 96.80 & 1.00 \\ 
 0.2 & 0.00 & 94.90 & 1.00 \\ 
 0.2 & 0.00 & 92.76 & 1.00 \\ 
 0.3 & 0.00 & 90.58 & 1.00 \\ 
 0.5 & 0.00 & 88.40 & 1.00 \\ \hline
 0.1 & 0.00 & 98.06 & 1.00 \\ 
 0.1 & 0.00 & 97.44 & 1.00 \\ 
 0.2 & 0.00 & 95.92 & 1.00 \\
 0.2 & 0.00 & 94.21 & 1.00 \\ 
 0.3 & 0.00 & 92.46 & 1.00 \\ 
 0.5 & 0.00 & 90.72 & 1.00 \\	\hline \hline	\end{tabular}
	\end{tabular}
\caption{Filters of different widths centered on 4000\AA\ are distorted by the presence of a non-telecentric angle, left for an F/1.4 and right for F/4}\label{table:squarefilternta}
\end{table}

\section{CONCLUSIONS}
 Narrow-band photometry is widely used to investigate stellar populations both with the Hubble Space Telescope\cite{Milone_2022} and with ground-based facilities\cite{Lee_2017}.
 %(see \citenum{Lee_2017, Milone_2022} and reference therein, for applications on star clusters).  
 The computer program presented in this paper will allow us to test the effectiveness of these magnitude determinations. Moreover, we will evaluate whether narrow-band filters can be successfully used in a fast wide-field telescope.

We can summarize the results of this study:
\begin{itemize}
    \item for each F-number exists a minimum bandwidth, a filter narrower than this will only decrease the efficiency of the system, not its line sensibility;
    \item even if the blue-shift effect of a small F-number can be taken into account in the design phase, a degradation of the transmission curve is unavoidable, we summarized it in the $I_{ratio}$ index;
    \item in non-telecentric optical designs a radial difference in the transmission curve must be considered. This can introduce a difference in photometry along the FoV. This can have an impact on new color indexes used in CMD or color-color diagrams. 
    \item there is monotonic blueshift with distance from the center of FoV; 
    \item there is a monotonic quality decrease of the transmission curves with distance from the center of FoV.
\end{itemize}

All these results must be taken into account when narrow-band photometry is foreseen in instruments with a very demanding optical design. A study of the behavior of a filter on a specific instrument can lead to a particular data reduction technique, e.g. considering rings of objects at the same distance from the center of the FoV for estimated corrections. 

\appendix

\section{Interference filters}\label{appen:optical}
%%%%%%%%%%%%%%%%%%%%%%%%%%%%%%%%%%%%%%%%%%%%%%%%%%%%%%%%%%%%%%%%%%%%%%%%%%%%%%%%%%%%%%%%%%%%%%%%%%%%%%%%%%%%%%%%%%%%%%
%%%%%%%%%%%%%%%%%%%%%%%%%%%%%%%%%%%%%%%%%%%%%%%%%% INTERFERENCE FILTERS %%%%%%%%%%%%%%%%%%%%%%%%%%%%%%%%%%%%%%%5%%%%%%
%%%%%%%%%%%%%%%%%%%%%%%%%%%%%%%%%%%%%%%%%%%%%%%%%%%%%%%%%%%%%%%%%%%%%%%%%%%%%%%%%%%%%%%%%%%%%%%%%%%%%%%%%%%%%%%%%%%%%%

%\input{FilterImage}

%\begin{figure}\label{figure:filtro e angolo} \caption{The effects of an interference filter, a certain (narrow) band is transmitted while the rest of the spectrum is reflected. As the distance from the normal increases, the filtered central $\lambda$ shifts to shorter wavelengths.}
%	\centering
%	\input{./TeX_files/Immaginefiltro2}
%\end{figure}
%\input{./TeX_files/Immaginefiltro2}

%\begin{itemize}
%	\item Filtro F-P, ricavare formule su trasmissione e sensibilità angolo \ref{math:lambdatheta}
%	\item film strati sottili $\lambda/4$ $\rightarrow$ filtro all-dielectric tipo HLHLHLSHLHLHL, seguengo \cite{Lissberger:59I} o \cite{macleod2010thin}
%	\item Presentare un po' di risultati sui filtri da \cite{macleod2010thin}
%	\item Posizionamento dei filtri nel telescopio, fascio convergente o fascio parallelo (perché non è la soluzione)
%	\item L'angolo non-telecentrico
%	\item Esigenze sul campo di vista
%\end{itemize}

In the current appendix, we will recollect the fundamental results regarding interference filters, the issues that emerge from the physics of these devices, and the characteristics that have to be considered for our purposes.  \\
As known, the transmission curve, the central wavelength in particular, of an interference filter is not always the same as the nominal one. Rather it changes relatively to the angle at which the radiation strikes the surface of the filter: in the case of a convergent beam, this means that different sections of the beam coming from the same source (arriving at different angles on the filter) find a filter whose properties are slightly different in front of it. The selection of wavelengths will be different from that nominally established and measured for a parallel beam with a propagation direction normal to the filter's plane.

\subsection[Interference filter F-P]{Interference filters \textit{a la} Fabry-Perot}

\begin{figure}
	\centering
	\begin{tikzpicture}[>=stealth]
		\draw(0,0)--(0,6);
		\draw (2,0)--(2,6);
		
		\draw[<->,dashed] (0,5)--(2,5) node[above,midway]{d};
		
		\draw[->] (-4,3) -- (-0.2,3)  node[above,midway]{$E_{in}$};
		\draw[->] (2.2,3) -- (6,3)  node[above,midway]{$E_{out}$};
		
		\node(a) at (1,-1) {semi-reflective surfaces};
		\fill[red] (a.north) circle (1pt);
		
		\draw [->] (0,3) node[above right]{$E_{in}t$} to[bend left=30] (2,3) ;
		\draw [->] (2,3) node[below left]{$E_{in}tr$} to[bend left=30] (0,3) ;
		
		\draw [->,dashed] (a.north) to[bend left=15] (2,1);
		\draw [->,dashed] (a.north) to[bend right=15] (0,1);
		
	\end{tikzpicture}
	\caption{Semi-reflective surfaces faced as in the Fabry-Perot interferometer. Incident radiation is partly transmitted and partly reflected.} \label{Figure:FP1}
\end{figure}
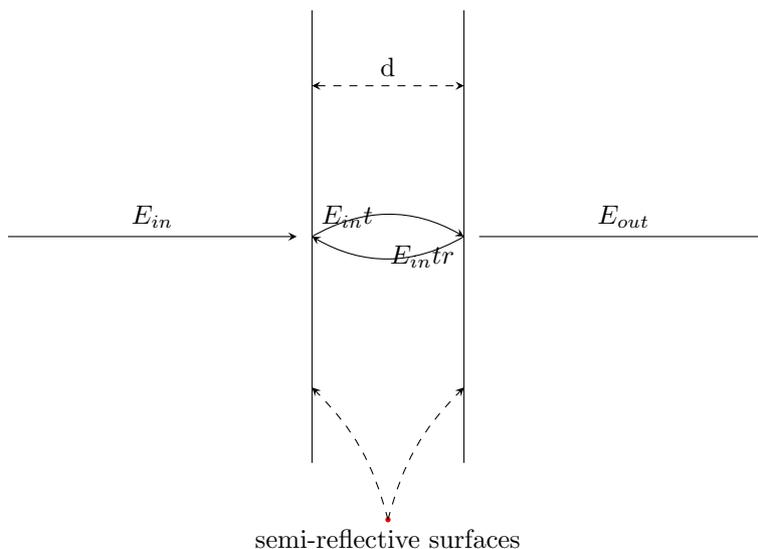

The simplest configuration for obtaining an interference filter is the one used by Charles Fabry and Alfred Perot in the late 19th century to construct their interferometer. Consisting of two identical partially reflecting surfaces parallel as in the figure \ref{Figure:FP1}.\\

%Consider an e.m. radiation, with an associated electric field $E_{in}$ propagating with direction perpendicular to the two surfaces. Where these have a transmission coefficient $T=t^2$, where $t$ is the transmission coefficient of the wave amplitude. The amplitude of the field just after the first mirror will be equal to $E_{in}t$, and as soon as the second reflecting surface is also crossed, we will have an amplitude equal to $$E_{out}^{(0)}=E_{in}t^2e^{i\frac{\phi}{2}} \label{eq:fp1}$$ where $\phi=\frac{2d}{\lambda}2\pi$ represents the phase shift occurred in the propagation. It is equal to twice the separation of the mirrors, i.e., equal to the round trip between the two. $\phi$ is sometimes called the \textit{roundtrip phase}. The radiation that is not transmitted by the second surface is reflected back to the first, which in turn reflects it back again always with the coefficient $R=r^2=1-T$. After these first two reflections $(E_{in}te^{i\frac{\phi}{2}})rre^{i\phi}$ we will have further transmission so that $$E_{out}^{(1)}=t^2E_{in}e^{i\frac{\phi}{2}}r^2e^{i\phi}=E_{out}^{(0)}r^2e^{i\phi}$$
%one can continue iteratively and consider that the amplitudes add up to 
%\begin{equation}\label{eq:fp2}
%	\frac{E_{out}}{E_{in}}=t^2e^{i\frac{\phi}{2}}(1+r^2e^{i\phi}+(r^2e^{i\phi})^2+(r^2e^{i\phi})^3+...)
%\end{equation}
%given that $r^2<1$ the series converges and the \ref{eq:fp2} becomes 
%\begin{equation}\label{eq:fp3}
%	\frac{E_{out}}{E_{in}}=\frac{t^2e^{i\frac{\phi}{2}}}{1-r^2e^{i\phi}}=\frac{(1-R)e^{i\frac{\phi}{2}}}{1-Re^{i\phi}}
%\end{equation}
%for which 

\begin{figure}
	\centering
	\begin{tikzpicture}[>=stealth]
		
		%%NORMALE
		\draw[dash dot,very thin] (-4,3) to (6,3);
		\coordinate (a) at (-4,2); \coordinate (b) at (0,3); \coordinate (c) at (-4,3); 
		\pic[draw,"$\theta$", draw=orange,<->,angle eccentricity=1.3,angle radius=3.5cm]{angle=c--b--a};
		
		\draw(0,0)--(0,6);
		\draw (2,0)--(2,6);
		
		\draw[<->,dotted] (0,1)--(2,1) node[above,midway]{d};
		
		\draw[->,thick] (-4,2) -- (0,3) node[above,midway]{$I_{in}$};
		\draw[dashed] (0,3) to (2,3.5);
		\draw[dashed] (2,3.5) to (0,4);
		\draw[dashed] (0,4) to (2,4.5);
		\draw[dashed] (2,4.5) to (0,5);
		\draw[dashed] (0,5) to (2,5.5);
		\draw[dashed] (2,5.5) to (1,5.75);

		\draw[->] (2,4.5) -- (6,5.5) node[above,midway]{$t''$} node[above]{$I_{t}$};
		\draw[->,dashed] (2,3.5) -- (6,4.5)  node[above,midway]{$t'$};
		\draw[->,dashed] (2,5.5) -- (6,6.5) node[above,midway]{$t'''$};
		
		\draw[->,dashed] (0,3) -- (-4,4) node[above,midway]{$r'$};
		\draw[->] (0,4) -- (-4,5) node[above,midway]{$r''$} node[above]{$I_{r}$};
		\draw[->,dashed] (0,5) -- (-4,6) node[above,midway]{$r'''$};
		
		\draw[dashed] (2,4.5) --(2.5,2.5);
		\draw[red] (2,3.5) to (2.23,3.555);
		
		\node[below left, red] at (0,3) {A};
		\node[below left, red] at (2,3.5) {B};
		\node[below left, red] at (0,4) {C};
		\node[below left, red] at (2,4.5) {D};
		\node[below right, red] at (2.23,3.555) {E};
	\end{tikzpicture}
	\caption{Non-perpendicular beam on a Fabry-Perot}\label{Figure:FP2}
\end{figure}
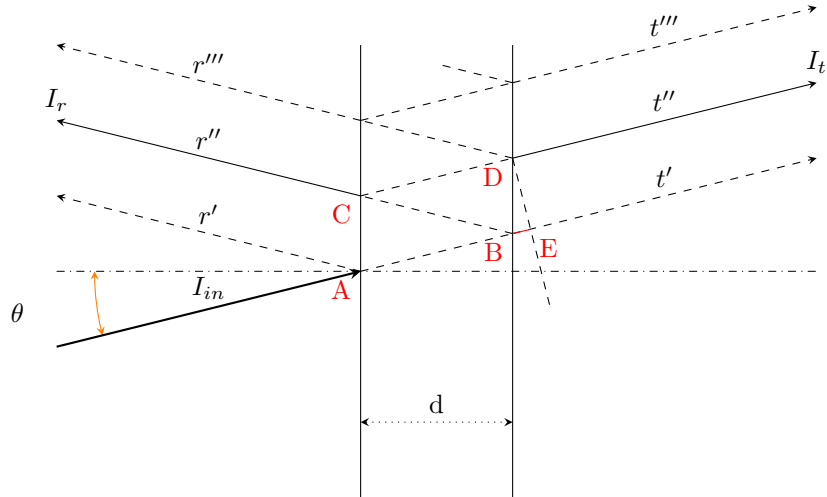

When these have a reflection coefficient $R=r^2$, where $r$ is the reflection coefficient of the wave amplitude, we have the maximum transmission 
\begin{equation}
	\left(\frac{E_{out}}{E_{in}}\right)_{max}=\frac{1-R}{1-R}=1
\end{equation}

when two times the distance between the two reflecting surfaces is equal to a multiple of the wavelength of the electromagnetic radiation. In other words, when we are dealing with standing waves between the two surfaces, the intensity of the transmitted radiation is equal to 100\% of the incident radiation.

\begin{equation}\label{eq:fbphi}
	\phi=n2\pi \Rightarrow \lambda=\frac{2d}{n}
\end{equation}
Conversely, for $\phi=n2\pi+\pi$ there is the minimum of transmission equal to $(\frac{E_{out}}{E_{in}})_{min}=\frac{1-R}{1+R}$. This minimum depends on $R$ so that the closer the reflectivity is to 1 the closer the transmission minimum is to 0. This is a typical interference effect that can be used to obtain a filter with a narrow bandwidth (which depends on R). \\
%So far, we have shown only the special case in which the radiation impinges perpendicularly on the mirrors. 
%It remains to find the relationship between the wavelength, which maximizes transmission, and the angle at which the radiation hits on the interferometer thus constituted. 
In Figure \ref{Figure:FP2} the radiation hits with an angle $\theta$ with respect to the normal: we see immediately that the optical path within the pair of mirrors (and hence the phase shift at the edges) changes with the same angle. We place ourselves here in the approximation of considering $d$ very small (of the same order as $\lambda$) so that the separation between successive beams $t',t'',...$ is such that interference is guaranteed. 
%Differently, one would have to place a convergent lens to bring the beams resulting from the successive reflections back to the interference condition\footnote{That is precisely the configuration of a classical F-P interferometer}.

As seen in the Figure \ref{Figure:FP2} the difference in the optical path between $t'$ and $t''$ is no longer simply equal to $2d$ but rather

\begin{equation}
	\begin{array}{l}
		t':AB+BE \\
		t':AB+BC+CD=3AB\\
		\Delta_{t''-t'}=2AB-BE
	\end{array}
\end{equation}

since $AB=\frac{d}{\cos\theta}$ and $$BE=BD\sin\theta=2AB=2d\frac{sin^2\theta}{\cos\theta}$$ then

\begin{equation}\label{eq:FPtheta}
	\Delta=2d\cos\theta
\end{equation}

From the Equation \ref{eq:fbphi}:
\begin{equation}
\lambda=\frac{2d\cos\theta}{n}
\end{equation}

The presence of an angular distance from the normal shifts the filter band toward the bluer wavelengths.\

Considering a thin foil\footnote{or otherwise a separating material with a different refractive index from that of the medium outside the two mirrors}, with a refractive index $\mu_f$, and interference due to multiple coherent beams, we obtain
\begin{equation}\label{eq:FPDeltathl}
\Delta=2\mu_f d\cos\theta
\end{equation}
and for the transmitted intensity $$I_t=I_i\textit{A}(\theta)$$ where the $\textit{A}(\theta)$ is the function of Airy $\textit{A}(\theta)=\frac{1}{1+F\sin^2(\phi/2)}$ where $F\equiv(\frac{2r}{1- r^2})^2$ is called the finesse coefficient and determines how narrow the transmission band is. One can consult Hecht\cite{Hecht} for details of the latter results.

\subsection{All-dielectric filters}\label{sec:alldielectric}

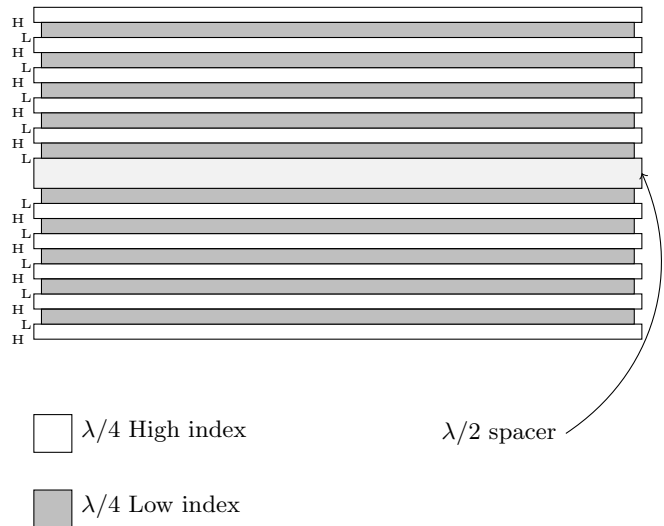
\begin{figure}
	\centering
	\begin{tikzpicture}
		\foreach \x in {0.,0.4,0.8,1.2,1.6}	{	
			\draw (0,0+\x) node[left]{\tiny H} rectangle (8,.2+\x);
			\draw[fill=black!25] (0.1,.2+\x)node[left]{\tiny L} rectangle (7.9,.4+\x);}
		\draw[fill=black!5](0,2) rectangle (8,2.4) ;
		\foreach \x in {0.,0.4,0.8,1.2,1.6}	{	
			\draw (0,2.6+\x) node[left]{\tiny H}  rectangle (8,2.8+\x);
			\draw[fill=black!25] (0.1,2.4+\x)node[left]{\tiny L} rectangle (7.9,2.6+\x);}
		
		\draw[->] (7,-1.25) node[left]{\footnotesize $\lambda/2$ spacer} to[bend right=40] (8,2.2);
		\draw (0,-1) rectangle (0.5,-1.5) node[above right]{\footnotesize $\lambda/4$ High index};
		\draw[fill=black!25] (0,-2.) rectangle (0.5,-2.5) node[above right]{\footnotesize $\lambda/4$ Low index};
	\end{tikzpicture}
	\caption{All-dielectrics filter}\label{figure:HL}
\end{figure}

An instrument built with two simple surfaces such as the Fabry-Perot interferometer leaves only one parameter free to determine the bandwidth of the filter and does not allow its shape to be changed. In particular, in order to obtain well-defined \textit {edge} filters and band-pass filters, it is essential to expand the reflecting surfaces into a stack of thin transparent and low-absorption dielectric layers. \\
An \textit {all-dieletrics} filter consists of a stack of dielectrics with a thickness of $\ lambda / 4 $ of refractive indexes of alternating values, a central resonant cavity, and another stack symmetrical to the first. In figure \ref{figure:HL}, an interference filter of this type is schematized, which also functions as a building block for more complex filters with two or three \textit{cavities}.

%Il calcolo della banda passante per un filtro così costruito prende le mosse da una serie di passaggi simili a quelli che conducono alla \ref{eq:FPDeltathl}, considerando lo spessore $\lambda/4$ e che riflessioni su interfacce da indice di rifrazione minore a maggiore comportano un cambiamento di fase $\pi$ mentre su interfacce da indice maggiore a minore non si ha cambiamento di fase\footnote{avevamo trascurato questo aspetto nella trattazione precedente perché in trasmissione tra sole due superfici le riflessioni sono sempre in numero pari}. Una teoria degli strati sottili con una trattazione completa dell'argomento e il suo svolgimento analitico-computazionale, che va oltre gli scopi del presente lavoro, si trova esposta in gran dettaglio nel \textcite{macleod2010thin} che inoltre contiene molti dati sperimentali e tecnico-industriali per la costruzione, il calcolo e la comprensione della teoria degli strati sottili sia per quanto riguarda i più semplici coating antiriflesso a singolo strato, sia per i filtri più complessi. \\
The first analytical treatment adequate to consider this type of filter with light that is not perfectly perpendicular is due to a series of works by P. H. Lissberger at the turn of 1960, in particular Lissberger\cite{Lissberger:59I} \cite{Lissberger:59II}. \\
In these works, Lissberger established the equivalence of an all-dielectric interference filter with a single-layer system (the spacer) with calculable reflectivity and with a transmission curve equal to the known Airy shape. Thus we obtain the wavelength of maximum transmission $ \lambda $ from $$ 2 \mu_ * d \cos \theta _ * = n \lambda $$ where $ \mu_* $ is the effective refractive index which takes a value intermediate between that of the H layers and that of the L layers. From here we get the \ref{math:lambdatheta} considering $ \theta_f $ the angle of propagation inside the equivalent filter, therefore for Snell's law from which, if the refractive index outside the filter is equal to 1, $ \sin \theta _ * = \frac {\sin \theta} {\mu_*} $.

\bibliography{Biblio.bib} % bibliography data in report.bib
\bibliographystyle{spiebib} % makes bibtex use spiebib.bst

%\bibliographystyle{alpha}
%\bibliography{sample}

%-------------------------------------REFERENCES
%\printbibliography[heading=bibintoc]

\end{document}